\documentclass[fleqn,usenatbib]{mnras}

\usepackage{newtxtext,newtxmath}
\usepackage{caption}
\usepackage{subcaption}
\usepackage{soul}
\usepackage{upgreek}

\usepackage{tikz}
\usetikzlibrary{shapes.geometric, arrows}

\tikzstyle{startstop} = [rectangle, rounded corners, minimum width=3cm, minimum height=1cm,text centered, draw=black, fill=white!30]
\tikzstyle{io} = [trapezium, trapezium left angle=70, trapezium right angle=110, minimum width=3cm, minimum height=1cm, text centered, draw=black, fill=white!30]

\tikzstyle{process} = [rectangle, minimum width=3cm, minimum height=1cm, text centered, draw=black, fill=white!30]

\tikzstyle{decision} = [diamond, minimum width=3cm, minimum height=1cm, text centered, draw=black, fill=white!30]

\tikzstyle{arrow} = [thick,->,>=stealth]

\usepackage[T1]{fontenc}

\DeclareRobustCommand{\VAN}[3]{#2}
\let\VANthebibliography\thebibliography
\def\thebibliography{\DeclareRobustCommand{\VAN}[3]{##3}\VANthebibliography}

\usepackage{graphicx}	
\usepackage{amsmath}	

\usepackage{amssymb}	

\let\oldhat\hat
\renewcommand{\hat}[1]{\oldhat{\boldsymbol{#1}}}
\newcommand{\Msun}{{\rm M}_{\odot}}
\newcommand{\mstar}{M_*}
\newcommand{\mdisc}{M_{\rm d}}

\newcommand{\Lsun}{{\rm L}_{\odot}}
\newcommand{\Rsun}{{\rm R}_{\odot}}

\newcommand{\Myr}{{\rm Myr}}
\newcommand{\mdot}{\dot{M}}
\newcommand{\Msunperyr}{{\,\rm M}_{\odot}\, {\rm yr}^{-1}}
\newcommand{\beq}{\begin{equation}}
\newcommand{\eeq}{\end{equation}}

\title[Semi-analytic model for episodic mass accretion]{A semi-analytic model for the temporal evolution of the episodic
disc-to-star accretion rate during star formation}

\author[Das \& Basu]{Indrani Das\thanks{E-mail:idas2@uwo.ca}
Shantanu Basu\thanks{E-mail:basu@uwo.ca}
\\
Department of Physics and Astronomy and Institute of Earth and Space Exploration, University of Western Ontario, London, Ontario N6A 3K7, Canada
}

\date{Accepted 2022 June 8. Received 2022 June 7; in original form 2021 December 24}

\pubyear{2021}

\begin{document}
\label{firstpage}
\pagerange{\pageref{firstpage}--\pageref{lastpage}}
\maketitle

\begin{abstract}
We develop a semi-analytic formalism for the determination of the evolution of the stellar mass accretion rate for specified density and velocity profiles that emerge from the runaway collapse of a prestellar cloud core. In the early phase, when the infall of matter from the surrounding envelope is substantial, the star accumulates mass primarily because of envelope-induced gravitational instability in a protostellar disc. In this phase, we model the envelope mass accretion rate from the isothermal free-fall collapse of a molecular cloud core. The disc gains mass from the envelope, and transports matter to the star via a disc accretion mechanism that includes episodic gravitational instability and mass accretion bursts according to the Toomre $Q$-criterion. 
In a later phase, mass is accreted on to the star due to gravitational torques within the spiral structures in the disc, in a manner that analytic theory suggests has a mass accretion rate $\propto t^{-6/5}$.
Our model provides a self-consistent evolution of the mass accretion rate by joining the spherical envelope accretion (dominant at the earlier stage) with the disc accretion (important at the later stage), and accounts for the presence of episodic accretion bursts at appropriate times. 
We show using a simple example that the burst mode 
can provide a good match to the observed distribution of bolometric luminosities. 
Our framework reproduces key elements of detailed numerical simulations of disc accretion and can aid in developing intuition about the basic physics as well as to compare theory with observations.

\end{abstract}




\section{Introduction} \label{sec:intro}
The luminosities of protostars from sub-millimetre/millimetre and mid-infrared observations of nearby molecular clouds reveal low to intermediate mass star formation occurring within dense protostellar cores. 
In the early phase (so-called Class 0 and Class I phases), a protostar is surrounded by a protostellar disc, which is in turn deeply embedded  within an infalling envelope that has emerged from the collapse of the surrounding cloud core. 
In order to achieve a typical final stellar mass within a certain timescale, the protostar accumulates mass with episodes of vigorous mass accretion rate ($\geq 10^{-4} \, \Msunperyr$) as observed in FU Orionis \citep[see][and references therein]{Audard_etal2014}. 
It is widely discussed that stellar mass is accumulated in a time-dependent and episodic (not steady) fashion. 
The sporadic mass accretion bursts can give rise to the luminous bursts observed in YSOs, which are often known as FUor ($\sim 100\, \Lsun$, $\sim 100\, {\rm yr}$ duration, thought to occur mainly in the embedded Class 0/I phase) and EXor ($\sim 10\, \Lsun$, $\sim 1 {\rm yr}$ duration, appearing in the Class II phase) eruptions \citep[see][references therein]{HartmannKenyon1996,Herbig2008, Audard_etal2014}.
The majority of the accretion outbursts that have been detected belong to the star-forming regions within the immediate solar neighbourhood. Less is known about episodic eruptions at larger distances ($\sim $ few kpc) due to drawbacks in time-domain astronomy at those distances.

An additional observational channel that can imply episodic accretion is the luminosity histogram of young stellar objects (YSOs). Several surveys show that the luminosities of YSOs are typically about an order of magnitude lower than that expected from a steady mass accretion rate \citep{Enoch_etal2009,Dunham_etal2010}. This has been used by the authors to provide further evidence of episodic accretion, in which the majority of time is spent in a lower accretion rate. However, long-term evolution of the accretion rate and low-amplitude variability can also potentially match the observed luminosities of protostars given favourable assumptions \citep{OffnerMckee2011,fischerEtal2017}. Further studies are required to distinguish these scenarios.

The overall protostellar accretion history can be thought of as a combined effect of two successive phases. 
In the early phase, when the infall of matter from the surrounding envelope is substantial, mass is transported inward because of envelope-induced gravitational instability in a protostellar disc. 
However, the accretion from the surrounding protostellar disc on to the central protostar primarily occurs at a much lower rate than the envelope accretion. 
In this early stage, even though both processes are coexisting, the envelope accretion dominates the disc accretion in driving the overall evolution. 
Simulations show that during the protostellar accretion phase (Class 0/I phase), intermittent accretion bursts occur due to episodes of dense clump infall on to the central protostar \citep{VB05L,vorobyov06}. Mass is infalling from the envelope to the disc, and the disc is transporting the matter to the star at a different rate. Because of the mismatch between the infall and transport rate, the central protostar mass does not grow at the same rate as the disc mass. Hence, the disc mass becomes comparable to that of the central protostar, which leads to gravitational instability (GI) within the protostellar/protoplanetary disc (PD/PPD). Generally, the disc gets fragmented into large spiral arms and gravitationally bound clumps.  Afterward, these clumps migrate inward through the spiral arms and fall to the centre \citep{VB05L,vorobyov06}, which triggers the luminosity bursts.

In the later phase, when the gas reservoir of the envelope is depleted, mass is
accreted on to the star due to internal torques, which usually result in a power-law decrease of mass accretion rate with time. Specifically, for accretion due to gravitational torques within the disc, analytic theory yields the form $\mdot \propto t^{-6/5}$ \citep[][hereafter LP]{LP87} that has also been found in numerical simulations \citep{vorobyov08}. At these times, disc accretion occurs independent of any exterior influence and the system moves into the Class II phase.

In this paper, we develop a semi-analytic formalism that successfully produces the evolution of the mass accretion rate for specified density and velocity profiles that emerge from the runaway collapse of a prestellar cloud core. 
We treat the prestellar core as an isothermal finite mass reservoir that is in hydrostatic equilibrium. In our model, we incorporate a prescription for generating the vigorous episodic outbursts that arise due to GI within the disc. The effect of GI can be characterized with the Toomre $Q$-parameter \citep{toomre1964}, which in turn can be expressed as a disc-to-star mass ratio that we use widely in our paper. We determine a self-consistent evolution of the mass accretion rate by joining the spherical envelope accretion and the disc accretion, together with the episodic accretion bursts. 

The semi-analytic model elucidates the basic physics of otherwise complex nonlinear simulations \citep[e.g.][]{VB05L,vorobyov06}. Aside from its simplicity and pedagogical value, it opens up new avenues for future work to model episodic accretion. The simulations are used to calibrate the model, and the computational efficiency of a model that requires mere seconds of wall clock run time opens up the possibility of running large parameter surveys and population synthesis modeling. We demonstrate a preliminary example of this by estimating a synthetic luminosity histogram in this paper. Such work can guide interpretation of whether the episodic accretion or a long timescale evolution of the accretion rate is the primary cause of the breadth of observed luminosity histograms. 

Our semi-analytic prescription for the determination of the mass accretion rate from the spherical collapse of an isothermal cloud core is described in Section \ref{sec:method}, where the two cases of spherical envelope accretion (Section \ref{sec:envelope_acc}) and disc accretion (Section \ref{sec:disc_acc}) are discussed separately. In Section \ref{sec:j_numerical} we also provide a semi-analytic prescription to determine the specific angular momentum profile. We describe the formulation of episodic outbursts due to GI in Section \ref{sec:bursts}. We present our numerical results in Section \ref{sec:results}. The logical flow of our calculation is described in Section \ref{sec:sa_pres}, the temporal evolution of the mass accretion rate is presented in Section \ref{sec:mdot_fig}, the modelling of the distribution of mass in the disc, star, and envelope is in Section \ref{sec:mass_fig}, and the estimation of the luminosity distribution is described in Section \ref{sec:lum_fig}. We discuss the applications and limitations of our results in Section \ref{sec:diss} and summarize the main features of our work in Section \ref{sec:sum}.

\section{Methodology}\label{sec:method}
We develop a semi-analytic prescription to determine the evolution of the protostellar mass accretion rate during star formation. 
This requires the modelling of three key ingredients: a prestellar density profile, the envelope accretion on to the disc, and the disc accretion. We describe these in the following subsections.
\subsection{Prestellar density profile}\label{sec:density}
We start with the modified isothermal density profile
\begin{equation}
    \rho(r) = \frac{\rho_c}{1+(r/r_c)^2} \ ,
\label{eq:rho_mis}    
\end{equation}
where $r_c$ and $\rho_c$ are the central lengthscale and density, respectively. 
The parameter $r_c$ represents the size of the central flat region of the density profile. Here, $r_c$ is comparable to the central Jeans length, $r_c = k c_s/\sqrt{\pi G \rho_c}$, such that inner profile remains close to that of a Bonnor-Ebert sphere, where $c_s$ is the isothermal sound speed and $G$ is the gravitational constant, $k$ ($=1.1$) is a constant of proportionality, $\rho_c = \mu m_{\rm H} n_c$ is the volume density, in which $\mu =2.33$ and $n_c$ is the central number density. The asymptotic density profile is $2.2$ times the equilibrium singular isothermal sphere value $\rho_{\rm SIS} = c_s^2/(2\pi G r^2)$ \citep[see more in][]{VB05mnras_a}. 
For the sake of convenience of our calculations, we normalize the physical quantities by the following units: 
$[L]= r_c$, $[t] = 1/\sqrt{G \rho_c}$, $[M] = \rho_c r_c^3$, $[\rho] = \rho_c$, $[v] = [L]/[t] = r_c 
\sqrt{G \rho_c}$, $[\dot{M}]=[M]/[t]$.
Our normalized variables are defined as $\tilde{r} = r/r_c$, $\tilde{t} = t/[t]$, $\tilde{M} =  M/[M]$, $\tilde{v} = v/[v]$, and $\dot{\tilde{M}} = \dot{M}/[\dot{M}]$. 

Next, we present the density profile of a tapered isothermal sphere,
\begin{equation}
    \rho(r) = \frac{\rho_c}{1+(r/r_c)^2} \left[1 - \frac{r^2}{R^2 _{\rm{out}}} \right] \ ,
\label{eq:rho_mod}     
\end{equation}
where $R_{\rm out}$ is the outer radius of the cloud. Note that we later use $R_{\rm out}$ as the key parameter to set the cloud mass. 
The density model of the tapered isothermal sphere provides a modified isothermal profile for an inner region, as well as a very steep outer density profile beyond some radial length scale. 
Qualitatively, this kind of transition in the density profile is consistent with the transition from an inner region with supercritical mass-to-flux ratio (gravitationally dominated) to a subcritical outer region (magnetically dominated) as found in numerical magnetohydrodynamic (MHD) simulations of the gravitational collapse of a a prestellar core \citep[e.g.][]{CM93, BM94}. 
It can also arise in hydrodynamic calculations due to a finite mass reservoir in the numerical domain \citep{VB05mnras_a}. 
The tapered isothermal sphere mimics these cases by having an enclosed mass that saturates to a final value as $r \rightarrow R_{\rm out}$.

\subsection{Spherical envelope accretion}\label{sec:envelope_acc}
We consider the collapse of the isothermal prestellar cloud core for modelling the spherical envelope accretion on to a central protostar and disc system. 
\subsubsection{Mass accretion rate from collapse of the cloud}
The equation of motion of a pressure-free, self-gravitating spherically
symmetric cloud is
\begin{equation}
\frac{dv}{dt} = -\frac{GM(r)}{r^2} \ ,
\label{eq:force_equn}
\end{equation}
where $v$ is the velocity of a thin spherical shell at a radial distance
$r$ from the centre of a cloud, and $M(r)$ is the mass inside a sphere
of radius $r$. Equation (\ref{eq:force_equn}) can be integrated to yield the expression for velocity $v(r, t)$ at a given radial distance $r$ and time $t$, which follows
\begin{equation}
    v = - \sqrt{2GM(r_0) \left(\frac{1}{r} - \frac{1}{r_0}\right)} \, ,
\label{eq:dim_vel} 
\end{equation}
{\rm where $r_0$ is the initial position of a mass shell at $t = t_0$}, and $M(r_0)$
is the mass inside $r_0$. Here, it is assumed that all shells are initially
at rest: $v_0(r_0) = 0$. Equation (\ref{eq:dim_vel}) can be integrated by 
introducing a new dimensionless variable $\beta$ such that
\begin{equation}
 r \equiv r_0 \cos^{2}\beta\, ,
\label{eq:beta} 
\end{equation}
\citep[see][]{Hunter1962}. Then the
time it takes for a shell located initially at $r_0$ to move to a smaller
radial distance $r$ due to the gravitational pull of the mass $M(r_0)$ is
\begin{equation}
    t = \frac{\arccos \sqrt{r/r_0} + 0.5 \sin(2\arccos \sqrt{r/r_0} \ \ )}{\sqrt{2GM/r_0^{3}}} \, .
\label{eq:dim_hunter}   
\end{equation}
The values of $r$ and $t$ are sufficient to determine $r_0$ (a value $>r$ but $<R_{\rm out}$, the radius of the cloud) from Equation (\ref{eq:dim_hunter}). 
The dimensionless form of Equations (\ref{eq:dim_vel}) and (\ref{eq:dim_hunter}) are
\begin{equation}
    \tilde{v} = - \sqrt{2\tilde{M}(\tilde{r}_0) \left(\frac{1}{\tilde{r}} - \frac{1}{\tilde{r}_0}\right)} \ ,
\label{eq:nondim_vel} 
\end{equation}
\begin{equation}
    \tilde{t} = \frac{\arccos \sqrt{\tilde{r}/\tilde{r}_0} + 0.5 \sin(2\arccos \sqrt{\tilde{r}/\tilde{r}_0})}{\sqrt{2\tilde{M}/\tilde{r}_0^3}} \ ,
\label{eq:nondim_hunter}     
\end{equation}
respectively, where $\tilde{r}_0 = r_0/r_c$.
Subsequently, we use the obtained value of $r_0$ in 
Equation (\ref{eq:nondim_vel}) to obtain $v(r, t)$.
Provided that the shells do not pass through each other (i.e. the
mass of a moving shell is conserved, so $dM(r, t) = dM(r_0, t_0$), the
gas density of a collapsing cloud is
\begin{equation}
    \rho(r, t) =  \rho_0 \>  \frac{r^{2}_0}{r^{2}} \> \frac{dr_0}{dr}  \ ,
\label{eq:rho}    
\end{equation}
where $\rho_0(r_0)$ is the initial gas density at $r_0$. The ratio of $dr_0/dr$
determines how the thickness of a given shell evolves with time.
The relative thickness $dr_0/dr$ is determined by differentiating Equation (\ref{eq:beta})
with respect to $r_0$, yielding
\begin{equation}
    \frac{dr}{dr_0} = \frac{r}{r_0} - r_0 \sin(2\arccos \sqrt{r/r_0}) \frac{d\beta}{dr_0} \, .
\label{eq:drdr0}    
\end{equation}
Next, $d\beta/dr_0$ is determined from an alternative form of Equation
(\ref{eq:dim_hunter}):
\begin{equation}
    \beta + \frac{1}{2} \sin 2\beta = t\, \sqrt{\frac{2GM(r_0)}{r_0^{3}}} \, ,
\label{eq:dim_beta}    
\end{equation}
and differentiating the above with respect to $r_0$ yields 
\begin{equation}
    \frac{d\beta}{dr_0} = \sqrt{\frac{G}{2M(r_0) r^{3}_0}} \>  \left(t\,\frac{r_0}{2r}\right) \left[\frac{dM(r_0)}{dr_0} - \frac{3M(r_0)}{r_0} \right] \ .
\label{eq:dim_dbetadr0}    
\end{equation}
Normalizing Equation (\ref{eq:dim_beta}) and Equation (\ref{eq:dim_dbetadr0}) we obtain
\begin{equation}
    \beta + \frac{1}{2} \sin 2\beta = \tilde{t}\, \sqrt{\frac{2\tilde{M}(\tilde{r}_0)}{\tilde{r}_0^3}} 
\label{eq:nondim_beta}      
\end{equation}
and 
\begin{equation}
    \frac{d\beta}{d\tilde{r}_0} = \sqrt{\frac{1}{2\tilde{M}(\tilde{r}_0) \tilde{r}^{3}_0}} \>  \left(\tilde{t}\,\frac{\tilde{r}_0}{2\tilde{r}}\right) \left[\frac{d\tilde{M}(r_0)}{d\tilde{r}_0} - \frac{3\tilde{M}(\tilde{r}_0)}{\tilde{r}_0} \right]  \ ,
\label{eq:nondim_dbetadr0}       
\end{equation}
respectively, where  
\begin{equation}
\frac{d\tilde{M}(\tilde{r}_0)}{d\tilde{r}_0} = 4\pi \tilde{\rho}_0(\tilde{r}_0) \tilde{r}^{2}_0 \, .
\label{eq:nondim_dMdr0}  
\end{equation}
Now that the density $\rho(r, t)$ and velocity $v(r,t)$ distributions of a
collapsing pressure-free sphere are explicitly determined, the mass
accretion rate at any given radial distance $r$ and time $t$ is 
\begin{equation}
    \dot{M}(r, t) = 4\pi r^2 \rho(r, t) v(r, t) \, .
\label{eq:mdot}    
\end{equation}
The normalized form of Equation (\ref{eq:mdot}) is
\begin{equation}
    \dot{\tilde{M}}(\tilde{r}, \tilde{t}) = 4\pi \tilde{r}^2 \tilde{\rho}(\tilde{r}, \tilde{t}) \tilde{v}(\tilde{r}, \tilde{t}) \, .
\label{eq:mdot_nondim}    
\end{equation}

\subsubsection{Evolution of mass accretion rate for different density models} \label{sec:evolmassacc}
We study the evolution of mass accretion rate for the spherical envelope accretion based on two different density models, a modified isothermal sphere and tapered isothermal sphere, given by Equations (\ref{eq:rho_mis}) and (\ref{eq:rho_mod}), respectively. The enclosed mass of the modified isothermal density model of Equation (\ref{eq:rho_mis}) is
\begin{equation}
    M(r_0) = 4\pi \rho_c r_c^2 \left(r_0 - r_c \arctan\frac{r_0}{r_c}\right) \, .
\label{eq:dimM_mis}    
\end{equation}
Normalization of Equation (\ref{eq:dimM_mis}) yields
\begin{equation}
    \tilde{M} (\tilde{r}_0) = 4\pi \left(\tilde{r}_0 - \arctan \tilde{r}_0 \right) \, .
\label{eq:nondimM_mis}    
\end{equation} 
Next, we calculate the enclosed mass for the tapered isothermal density model of Equation (\ref{eq:rho_mod}), which yields
\begin{equation}
\begin{aligned}
    M(r_0) = 4\pi \rho_c r_c^3 & \int_0 ^{r_0/r_c} \frac{\tilde{r}^2}{1+\tilde{r}^2} d\tilde{r}  \\
                & - \frac{4\pi \rho_c r_c^5}{R^2_{\rm{out}}}  \int_0 ^{r_0/r_c} \frac{\tilde{r}^4}{1+\tilde{r}^2} d\tilde{r} \ ,
\end{aligned}
\label{eq:dimM_mod_int}     
\end{equation}
where 
$\tilde{r}$ is the dimensionless radial lengthscale described in Section \ref{sec:density}. After integrating Equation (\ref{eq:dimM_mod_int}) we obtain
\begin{equation}
\begin{aligned}
    M(r_0) = 4\pi \rho_c r_c^2 & \left(r_0 - r_c \arctan \frac{r_0}{r_c} \right) \\
                               & - \frac{4\pi \rho_c r_c^5}{R^{2}_{\rm{out}}} \left(\arctan \frac{r_0}{r_c} + \frac{r^{3}_0}{3 r_c^{3}} - \frac{r_0}{r_c} \right)  \ .
\end{aligned}
\label{eq:dimM_mod}     
\end{equation}
Normalizing Equation (\ref{eq:dimM_mod}) yields
\begin{equation}
\begin{aligned}
    \tilde{M} (\tilde{r}_0) = 4\pi \left[ \left(\tilde{r}_0 - \arctan \tilde{r}_0 \right) - \frac{1}{\tilde{R}^{2}_{\rm{out}}} \left(\arctan \tilde{r}_0 + \frac{\tilde{r}^{3}_0}{3} - \tilde{r}_0 \right) \right] \ .
\end{aligned}    
\label{eq:nondimM_mod}     
\end{equation}

\subsection{Disc accretion} \label{sec:disc_acc} 
The disc accretion plays an important role in the formation and evolution of disc. During the early stellar evolution the solar nebula was an accretion disc \citep[see discussion in][]{Pringle1981}. Primarily, the matter falls on to the disc from the envelope and then it is partially transported to the star from the disc. In a simple viscous disc, matter is transported inward in an axisymmetric model and results in an accretion rate on to the central star. At the same time, the disc evolution results in a loss of angular momentum associated with the accreting material \citep{LyndenBell_Pringle1974, Pringle1981}. \cite{HartmannLee_etal98} applied such a model in which the accretion and angular momentum transfer/redistribution occurs due to a well-defined viscosity $\nu \propto R^{\gamma}$, where $R$ is the disc radius. This viscosity profile yields a declining mass accretion rate $\dot{M}_{\rm disc} \propto t^{-\eta}$, where $\eta \gtrsim 1.5$ corresponds to $\gamma \gtrsim 1$. Note that any $\eta > 1$ corresponds to a finite mass reservoir. 
\cite{HartmannLee_etal98} discussed that the limit $\gamma \sim 1$ essentially corresponds to roughly constant $\alpha$ in the turbulent viscosity parameterization of \cite{ShakuraSunyaev1973}, which is $\nu = \alpha c_s H$, where $H$ is the disc scale height (half thickness).

In a more realistic disc, the gravitational torques produced by the nonaxisymmetric structure including spiral arms can dominate the angular momentum transfer.
To model such evolution we use the gravitational torque driven accretion \citep{LP87}, which leads to $\dot{M}_{\rm disc} \propto t^{-6/5}$, which was also found in simulations by \citet[see their fig. 3]{vorobyov08} for late time evolution when envelope accretion had become negligible. The mass of the disc during its evolution can be written as
\begin{equation}
    \mdisc (t) = \frac{4\pi}{3} \Sigma_{0d} R_{0d}^2 \left(\frac{t}{t_{0d}}\right)^{-1/5} \ ,
\label{eq:mass_LP}
\end{equation}
as described in \cite{LP87}, where $M_{0d} (t=t_{0d}) \equiv (4\pi/3) \Sigma_{0d} R_{0d}^2$. 
So, taking the time derivative of Equation (\ref{eq:mass_LP}) yields the mass accretion rate from the disc to the star as
\begin{equation}
    \dot{M}_{\rm ds}(t) = - \frac{1}{5} \left(\frac{4\pi}{3} \Sigma_{0d}\, R_{0d}^2 \,t_{0d}^{1/5} \right) \> t^{-6/5} \ .
\label{eq:mdotLP}    
\end{equation}
Normalization of Equation (\ref{eq:mdotLP}) yields
\begin{equation}
    \dot{\tilde{M}}_{\rm ds}(\tilde{t}) = - \frac{C_1}{5} \tilde{t}^{-6/5} \ ,
\label{eq:nondim_mdotLP}    
\end{equation}
where 
\begin{equation}
    C_1 =  \frac{\frac{4\pi}{3} \Sigma_{0d} R_{0d} ^2}{\rho_c r_c^3} \left(\frac{t_{0d}}{[t]} \right)^{1/5} \equiv \frac{M_{0d}}{[M]} \left(\frac{t_{0d}}{[t]} \right)^{1/5} = \tilde{M}_{0d} \tilde{t}^{1/5}_{0d} \ ,
\label{eq:C1}    
\end{equation}
and $\dot{\tilde{M}}_{\rm ds}(\tilde{t}) = \dot{M}_{\rm ds}(t)/[\dot{M}]$. Here $[\dot{M}]$, $[t]$, and $[M]$ are defined in the previous section. Note that use of $C_1$ requires us to initialize the disc evolution with an initial mass $M_{0d}$ at time $t_{0d}$. These values are episodically updated in a manner that depends on the mass accretion bursts and is described in Section \ref{sec:bursts}.

We take note of the fact that all accretion on to the star is not occurring purely through the disc. Our simplified spherical envelope accretion model implies that some matter would accrete in a direction along or nearly along the rotation axis. Even in a more realistic geometry, some infall may be funneled along the edge of the outflow cavity and directly reach the innermost region of the stellar magnetosphere. Furthermore, the disc-to-star accretion may itself be enhanced above the values calculated above for an isolated disc when there is forcing from envelope accretion. For all of these reasons, we adopt the total accretion rate of mass reaching the star to be
\begin{equation}
   \dot{M}_* =  \dot{M}_{\rm ds}(t)  + q\,\dot{M}_{\rm infall}(t),
   \label{eq:dotmstar}
\end{equation}
where $q$ is the fraction of envelope mass accretion that goes directly
to the star and $\dot{M}_{\rm infall}$ is the accretion rate calculated according to the prescription in Equation (\ref{eq:mdot}). We adopt $q=0.1$, which means in practice that the second term in Equation (\ref{eq:dotmstar}) will dominate the first until $\dot{M}_{\rm infall}$ undergoes a rapid drop due to the tapering of the isothermal density profile. Once this has happened, the $\dot{M}_*(t)$ will equal $\dot{M}_{\rm ds}(t)$ for subsequent evolution.



\subsection{Episodic accretion bursts} \label{sec:bursts}
We invoke the condition for the occurrence of episodic accretion bursts in terms of the Toomre-$Q$ instability criterion:
\begin{equation}
Q \equiv \frac{c_s \Omega}{\pi G \Sigma} \lesssim 1 \, ,
\label{eq:ToomreQ}    
\end{equation}
where $\Omega = \sqrt{GM_*/R_{\rm d}^{3}}$ is the Keplerian angular speed, $\Sigma$ is the disc surface density, $M_*$ is the stellar mass, $\mdisc$ ($\equiv \pi R_{\rm d}^2 \Sigma$) is the disc mass, $R_{\rm d}$ is the disc radius, and $c_s$ and $G$ are defined earlier. We use $c_s/ \Omega = H$ for a non-self-gravitating disc in hydrostatic balance, where $H$ is the vertical half-thickness of the disc. Rearranging the parameters within Equation (\ref{eq:ToomreQ}) yields 
\begin{equation}
    Q \simeq \frac{H}{R_{\rm d}} \frac{M_*}{\mdisc} \ .
\label{eq:ToomreQ_ds_ratio}    
\end{equation}
Therefore, the Toomre instability criterion can be written as
\begin{equation}
    \frac{\mdisc}{M_*} \gtrsim \frac{H}{R_{\rm d}} \ . 
\label{eq:ds_ratio_HR}    
\end{equation}
For a typical disc, $H/R_{\rm d}$ is a few times 0.1 \citep[see discussion in][]{kratterLodato2016}. 
The factor $H/R_{\rm d}$ also includes the effects of the disc surface density profile.
The Equation (\ref{eq:ds_ratio_HR}) is another way of representing gravitational instability in terms of disc-to-star-mass ratio, which is a constraint that is more related to observations. In our model, this depiction is widely used.
The disc becomes gravitationally unstable, and a burst occurs, if the disc-to-star-mass ratio satisfies
\begin{equation}
    \frac{\mdisc}{M_*} \gtrsim \mathcal{R}_{\rm b} \, ,
\label{eq:ds_ratio_rp}      
\end{equation}
where we consider $\mathcal{R}_{\rm b}=0.33$ as found in the hydrodynamic simulations \citep[e.g.][]{vorobyov06}. 
The value of $\mathcal{R}_{\rm b}$ can vary up to $10\%-20\%$ depending on the numerical model.
The disc becomes gravitationally stable after the burst, which means the updated disc-to-star-mass ratio ($\mathcal{R}_{\rm f}$) becomes fairly lower than the threshold. However, it can still be of the order few times $\sim 0.1$. 
A very small $\mathcal{R}_{\rm f}$ ($\lesssim 0.1$) is not a good choice as it produces massive bursts that could exceed typical values ($\sim 0.01 -0.05\, \Msun$) by a factor up to $10-15$.
There is some freedom in choosing $\mathcal{R}_{\rm f}$ and we set it to $0.23$ in order to obtain a reasonable number of bursts in comparison to simulations. A higher value will result in more bursts and a lesser value yields fewer bursts.
During a burst, the disc loses mass and the star gains mass, both in an amount equal to the mass of the clump. So, the updated disc-to star mass ratio becomes
\begin{equation}
    \frac{M^\dagger_{\rm d}}{M^\dagger_*} = \mathcal{R}_{\rm f} \, , 
\label{eq:ds_ratio_rpd}      
\end{equation}
where $\mathcal{R}_{\rm f} < \mathcal{R}_{\rm b}$, $M^\dagger_{\rm d} = \mdisc -M_{\rm burst}$, and $M^\dagger_* = M_* + M_{\rm burst}$. 
So, simplifying Equation (\ref{eq:ds_ratio_rpd}) the clump mass (often called as burst mass) can be calculated as
\begin{equation}
    M_{\rm burst} = \frac{\mdisc - \mathcal{R}_{\rm f} \,M_*}{(1+\mathcal{R}_{\rm f})} \ .
\label{bm_eq}    
\end{equation}
After every burst, we modify $M_{0d}$ with the updated disc mass. We further calculate $\dot{M}_{\rm ds}$ and $C_1$ as shown in Equation (\ref{eq:nondim_mdotLP}) and  (\ref{eq:C1}) using the updated disc mass until the 
next burst. We describe our numerical prescription in detail in Section \ref{sec:sa_pres}.
In our formalism we set the time duration for each burst as
\begin{equation}
   \Delta t_{\rm burst} =100 \ \frac{M_{\rm burst}}{0.01\, \Msun} \  {\rm yr} \ .
\label{eq:deltat_burst}  
\end{equation} 
For simplicity, we correlate the time duration of the burst with the mass of the burst in a linear fashion such that time duration of an episodic burst of typical mass $0.01\, \Msun$ is around $100 \ {\rm yr}$ as found in the simulations and observations \citep[][references therein]{HartmannKenyon1996, Herbig2008, Audard_etal2014}. 
The linear relation between the burst mass and its time duration in Equation (\ref{eq:deltat_burst}) results in an increasing burst duration with time since the relatively more massive clumps (few times $0.01\, \Msun$) that form at later times in the more massive discs will be tidally sheared into multiple clumps of $\sim 0.01\, \Msun$ that collectively produce a longer duration burst \citep{vor15}.  
A linear relation is the simplest function that can be used to show an increasing duration as the burst masses become larger in the later evolution. 
The simulations do not provide a large number of data points in this regard, so a linear relation is the simplest fitting function and exhibits a luminosity evolution that is reasonable in comparison to the simulations. 



\section{Results} \label{sec:results}
We implement our formalism for models with three different parent core masses as described in Table \ref{tab:model}.
\begin{table*}
\caption{Model parameters: temperature $T$, the central number density $n_c$, the central flat region of the core $r_c$, outer radius of the core $R_{\rm out}$, mass of core $M_{\rm core}$, and final stellar mass $M_*$ at the end of the desired evolution.}
\begin{center}
\label{tab:model}
\begin{tabular}{c c c c c c c c}
    \hline
    & $T$ & $n_c$ & $r_c$ & $R_{\rm out}$ & $R_{\rm out}/r_c$ &  $M_{\rm core}$ & $M_*$ \\
    & [K] &  [$\times\, 10^4 \ {\rm cm}^{-3}$] & [pc] & [pc] & & [$M_{\odot}$] & [$M_{\odot}$]\\
    \hline \\
    \textsc{model1} & 12 & 4.95 & 0.037 & 0.118 & 3.2 &  0.50 & 0.28\\
    \hline
     \textsc{model2} & 12 & 4.95 & 0.037 & 0.148 & 4.0 & 1.10 & 0.57\\
    \hline
    \textsc{model3} & 12 & 4.95 & 0.037 & 0.185 & 5.0 & 1.98 & 1.0\\
    \hline
    \textsc{model2a} & 12 & 8.5 & 0.028 & 0.112 & 4.0 & 0.84 & 0.45\\
    \textsc{model2b} & 12 & 2.0 & 0.058 & 0.232 & 4.0 & 1.74 & 0.9\\
    \hline
    \textsc{model2c} & 8 & 3.25 & 0.037 & 0.148 & 4.0 & 0.75 & 0.39\\
    \textsc{model2d} & 16 & 6.70 & 0.037 & 0.146 & 4.0 & 1.46 & 0.77\\
    \hline
\end{tabular}
\end{center}
\end{table*} 
We study cores of different masses by varying the outer radius. 
The actual size of the prestellar cloud core is one of the significant constraints from observations. For example, 
in the dense cores in Taurus,
$0.02\, {\rm pc} \lesssim R_{\rm edge} \lesssim 0.1\, {\rm pc}$
\citep[e.g.][]{motte2001}. Simulations show that beyond some such distance $R_{\rm edge}$, the column density merges into a near-uniform background or fluctuates about a typical mean value resembling the ambient molecular cloud \citep[][]{basu04}.

\subsection{Envelope accretion}\label{sec:env_accre}
\begin{figure}
	\includegraphics[width=\columnwidth]{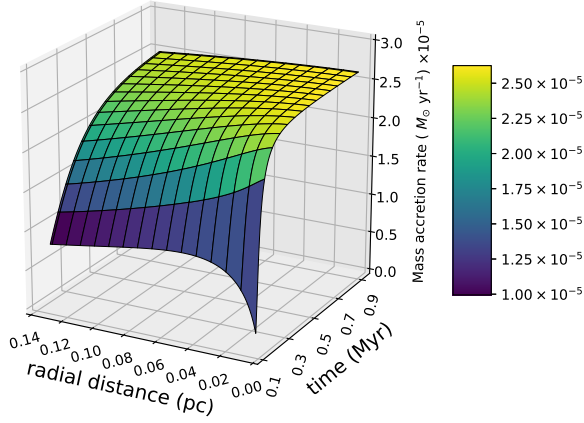}
    \caption{2D representations of mass accretion rate as a function of space and time for the isothermal density profile.}
    \label{fig:mdot_const}
\end{figure}

We solve the mass accretion rate using Equation (\ref{eq:mdot}) for the modified isothermal density profile of Equation (\ref{eq:rho_mis}) by combining the Equations (\ref{eq:force_equn})--(\ref{eq:mdot}). 
Figure \ref{fig:mdot_const} shows the mass accretion rate $\mdot(r,t)$ as a function of space and time for a pressure-free cloud with $n_c \sim 5 \times 10^4 \ {\rm cm}^{-3}$ and $r_c=0.037 \ {\rm pc}$. 
\begin{figure}
	\includegraphics[width=\columnwidth]{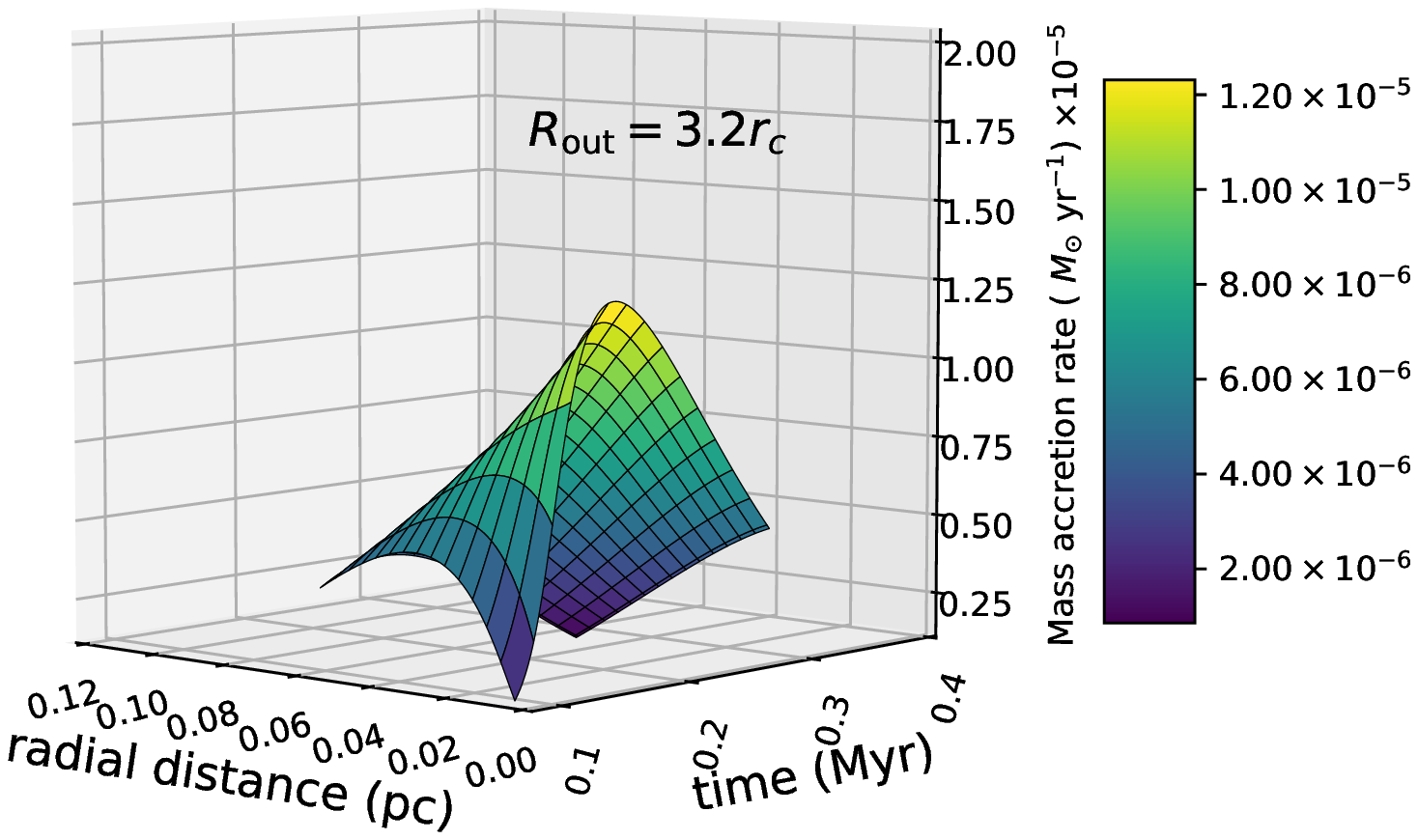}{(a)}
	\includegraphics[width=\columnwidth]{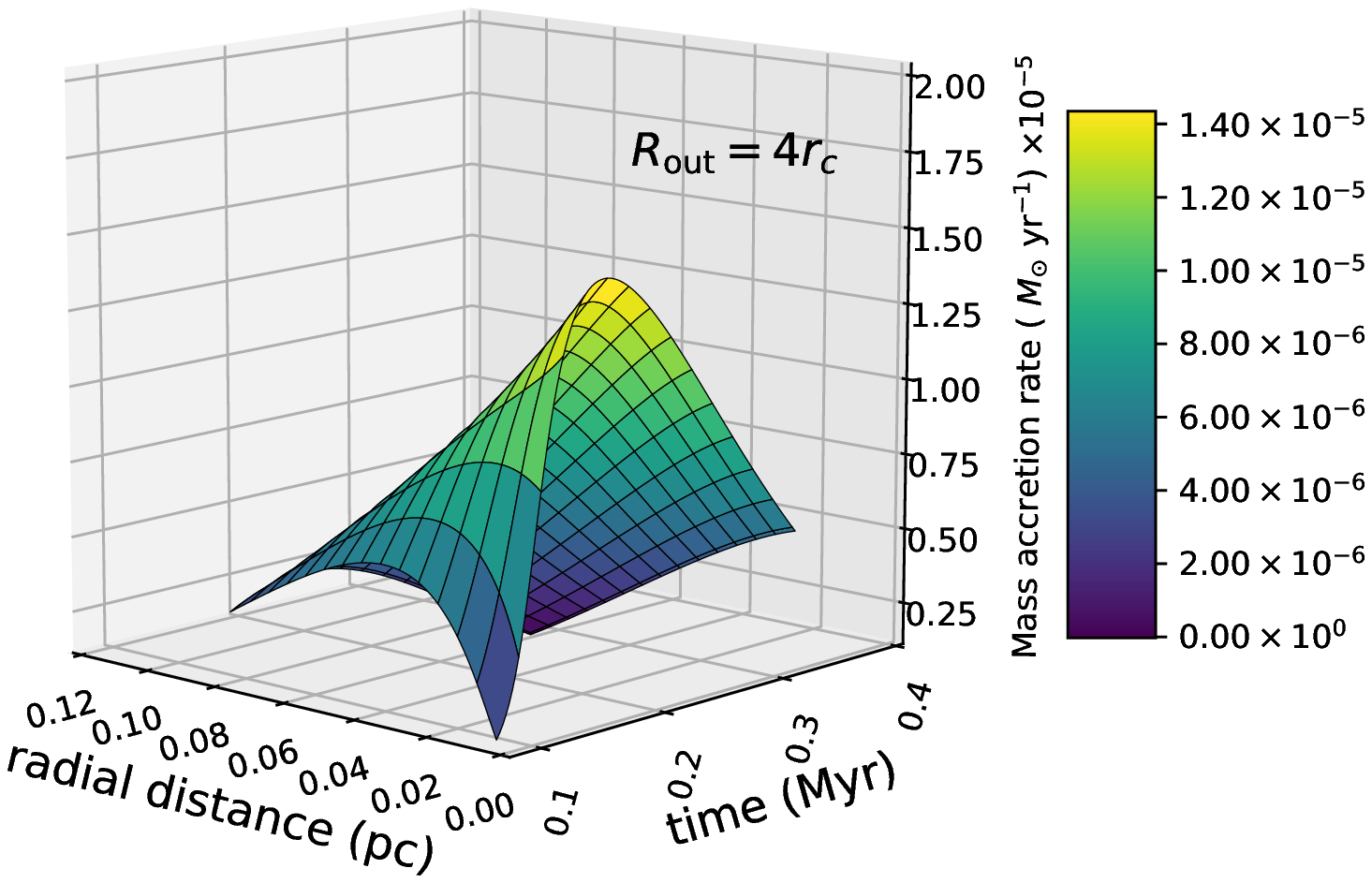}{(b)}
	\includegraphics[width=\columnwidth]{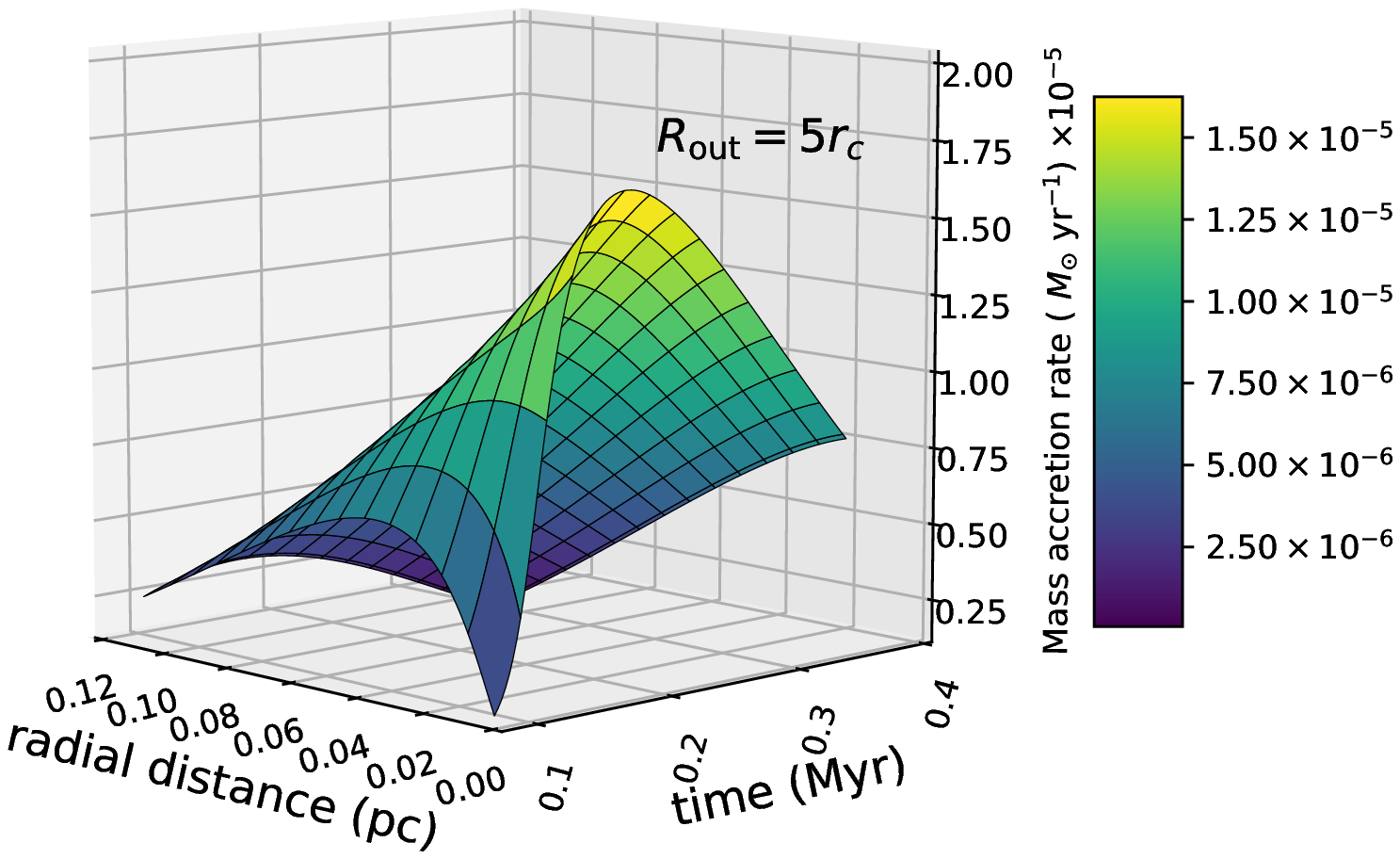}{(c)}
    \caption{2D representations of mass accretion rate as a function of space and time for the modified density profile with different outer radius (a) $R_{\rm out} = 3.2 r_c$, (b) $R_{\rm out} = 4 r_c$, (c) $R_{\rm out} = 5 r_c$.}
    \label{fig:mdot_Rout_3D}
\end{figure}
The mass accretion rate initially increases with time and appears to approach a constant value at later times $t > 0.7 \ \Myr$. The temporal evolution of the mass accretion rate also has a radial dependence. Hence, we see that at smaller radial distances, $\mdot(r,t)$ approaches the constant value at a faster time. 
In Figure \ref{fig:mdot_Rout_3D}a, b, c  we present the mass accretion rate $\mdot (r,t)$ as a function of space and time for the tapered isothermal density profile of Equation (\ref{eq:rho_mod}) with three different outer radii $R_{\rm out} = 3.2 r_c, \ 4r_c, \ 5r_c$, respectively. Initially $\mdot (r,t)$ increases with time. Afterwards, because of the finite mass reservoir when the envelope is depleted, $\mdot (r,t)$ gradually decreases over the time as well as over the radial distance. We find the $\mdot (r,t)$ rises faster to a peak at a smaller radial lengthscale. With increasing $R_{\rm out}$, $\mdot (r,t)$ falls off at a later time.
\begin{figure}
\includegraphics[width=\columnwidth]{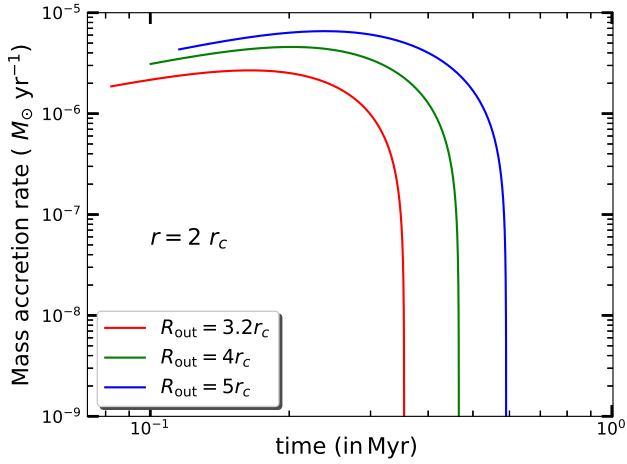}
\caption{1D representation of the temporal evolution of mass accretion rate for models with $R_{\rm out} = 3.2 r_c$ (red line), $R_{\rm out} = 4 r_c$ (green line), $R_{\rm out} = 5 r_c$ (blue line) calculated at a fixed radial distance $r=2 r_c$.}
\label{fig:mdot_Rout_2D}
\end{figure}
Figure \ref{fig:mdot_Rout_2D} shows the temporal evolution of $\mdot (r,t)$ at a fixed radial distance $r=2 r_c$ for all three models as shown in Figure \ref{fig:mdot_Rout_3D}. It is evident from this plot that, with increasing $R_{\rm out}$, $\mdot$ starts to evolve at a slightly later time and attains the maximum and then falls after a longer time.

\begin{figure}
\includegraphics[width=0.95\linewidth]{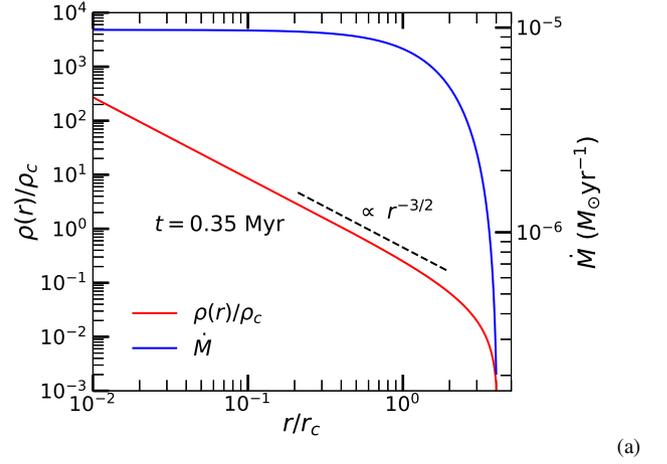}{(a)}
\includegraphics[width=0.95\linewidth]{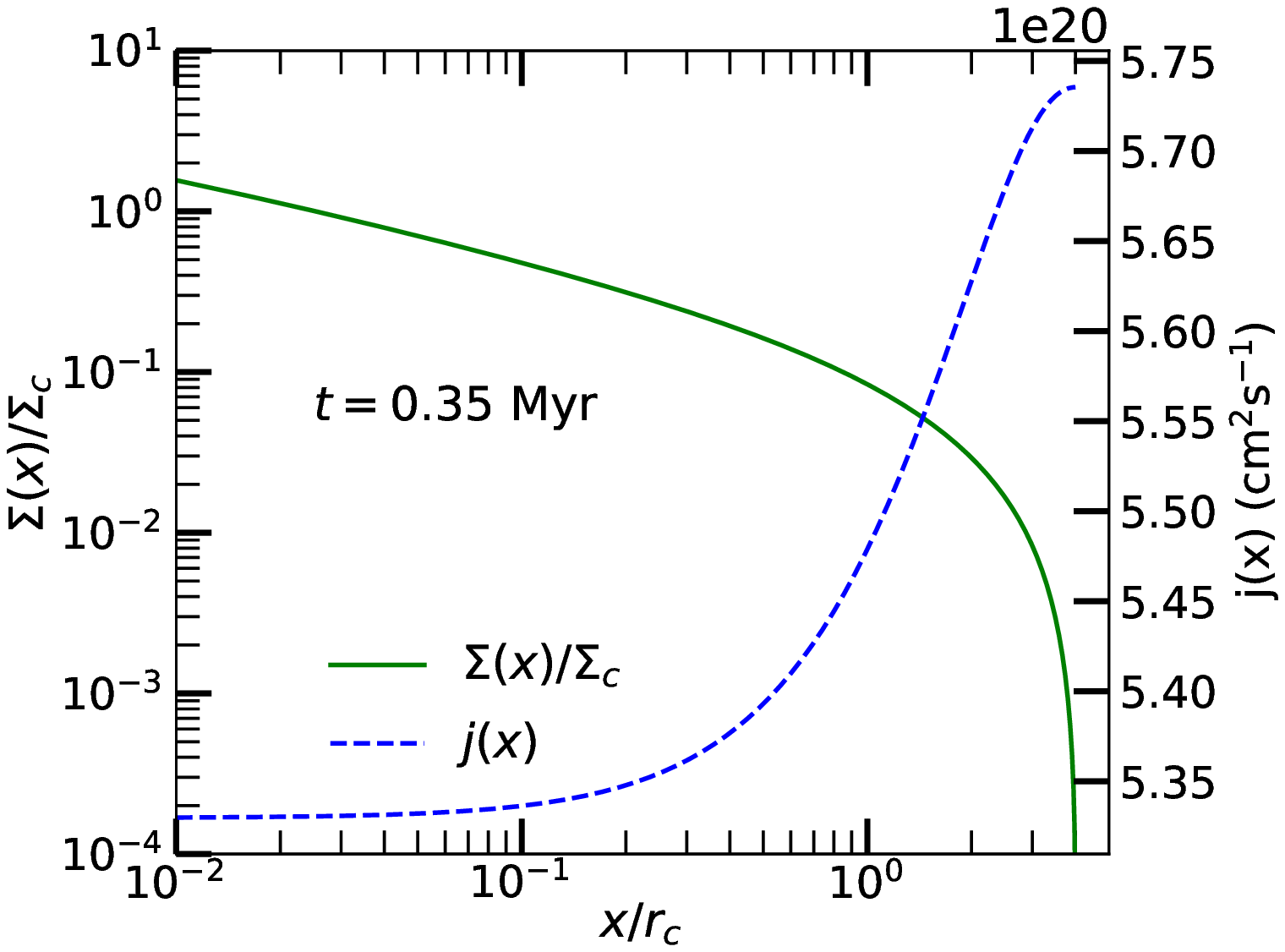}{(b)}
\caption{Top: Radial profiles of density $\rho(r)/r_c$ (red) and mass accretion rate $\dot{M}$ (blue) as a function of radial distance $x/r_c$ at a fixed time $t=0.35 \ {\rm Myr}$ for model with $R_{\rm out} = 5 r_c$. Bottom: Column density $\Sigma(x)/\Sigma_c$ (green) and the specific angular momentum $j(x)$ (blue) as a function of radial offset $x/r_c$ at the same time for the same model. 
}
\label{fig:Rout5_2D_radial_jMSigma}
\end{figure}

\subsubsection{Spatial profiles of mass and angular momentum} \label{sec:j_numerical}
\begin{figure}
\centering
\includegraphics[width=0.85\linewidth]{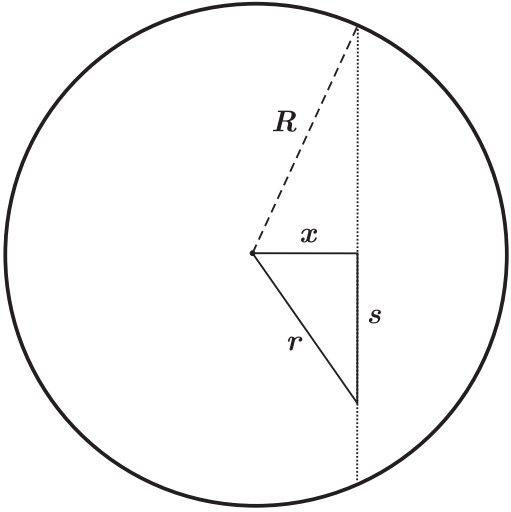}{}
\caption{Schematic illustration of a cut through a spherical cloud of radius $R$. The observer is positioned along the direction of the coordinate $s$, and measures an integrated column density $\Sigma$ as a function of the offset $x$ \citep[figure taken from][]{dappBasu09}.}
\label{fig:geometry}
\end{figure}

In our model we deal with the geometry of a spherical cloud with an outer radius $R_{\rm out}$. To obtain the column density and angular momentum as function of an observationally tractable parameter, we consider a cut through of the spherical cloud. If we are positioned along the direction of line-of-sight coordinate $s$, then we can measure these quantities as function of radial offset $x$ using the transformation from $s$ coordinate to the radial offset coordinate $x$, which yields $s = \sqrt{r^2-x^2}$ and hence $ds = rdr/\sqrt{r^2- x^2}$. 
The geometry is portrayed in Figure \ref{fig:geometry}. 
The column density can be calculated by integrating the volume density
along a line of sight through the spherical cloud:
\begin{equation}
\begin{aligned}
    \Sigma(x) & = 2 \int_0^{\sqrt{R_{\rm out}^2-x^2}} \rho(s) ds  \\
              & =  \int_0^{R_{\rm out}} \frac{\rho(r) r}{\sqrt{r^2 - x^2}} dr\, ,
\end{aligned} 
\label{eq:sigma_x_formula}
\end{equation}
as described in \S\ 2.1 of \cite{dappBasu09}. 
Once $\Sigma(x)$ is obtained, we calculate the enclosed mass by integrating the column density along the radial offset through the sphere, yielding
\begin{equation}
    M(x) = \int_0 ^x \Sigma(x^{\prime})\  x^{\prime} dx^{\prime} \ .
\label{eq:M_x_formula}
\end{equation}

In Figure \ref{fig:Rout5_2D_radial_jMSigma}a, we show the evolution of the density $\rho(r)/\rho_c$ (red line) as a function of radial distance $r/r_c$ at a later time $t= 0.35 \ {\rm Myr}$. For $r/r_c \lesssim 1$, $\rho(r) \propto r^{-3/2}$, characteristic of an expansion wave \citep[][]{Shu1977}; and for $r/r_c > 1$, $\rho(r)$ starts to diverge and sharply falls. In our model, $R_{\rm out}$ is not large enough to show a clear power law $\rho(r) \propto r^{-2}$ profile before the steep descent near the edge. 
In Figure \ref{fig:Rout5_2D_radial_jMSigma}a, the blue line presents the mass accretion rate $\dot{M}$ (blue line) as a function of $r/r_c$ at $t = 0.35 \ {\rm Myr}$. In our work, we exclude the case of large $R_{\rm out}/r_c \sim 10 \  {\rm or} \ 100$ as it will give rise to intermediate to high mass stars.

In Figure \ref{fig:Rout5_2D_radial_jMSigma}b, the green line shows $\Sigma(x)/\Sigma_c$ 
as a function radial offset $x/r_c$ at a time $t=0.35 \ {\rm Myr}$ for the model with $R_{\rm out} = 5r_c$. Here, $\Sigma_c = \Sigma(x=0) = 2 r_c \rho_c \ {\rm arctan}(R_{\rm out}/r_c)$ for the modified isothermal density profile shown in Equation (\ref{eq:rho_mis}) \citep[see more in][]{dappBasu09}. 
The profile of $\Sigma(x)/\Sigma_c$ essentially traces the profile of $\rho(r)/\rho_c$ and falls sharply at the outer edge. Hence, $M(x)$ gets saturated at the outer radius. 
However, the conservation of the specific angular momentum corresponds to the total enclosed mass $M_{\rm tot}(x)$. It can be calculated as $M_{\rm tot}(x) = M(x) + M_*(t)$, which includes the mass that has reached the star. Here, $M_*(t)$ is the stellar mass at the centre at time $t$. 
The relation between specific angular momentum $j(x)$ and the total enclosed mass $M_{\rm tot}(x)$ can be written as 
\begin{equation}
    j(x) = \frac{\Omega_0 \pi G^{1/2}}{B_{\rm ref}} \ M_{\rm tot}(x)\, ,
\label{eq:j_x_formula}
\end{equation}
as shown in \cite{Basu98}. Here, $\Omega_0 $ is the typical (mean) rotational rate for the dense molecular cloud and is about $10^{-14} \ {\rm rad \  s^{-1}}$ \citep[see][]{Goodmanetal1993}, and $B_{\rm ref}$ is the canonical magnetic field of $30 \ \upmu G$ for a molecular cloud core \citep[see][]{Goodmanetal1993,Crutcher1993ApJ}. We follow a semi-analytic approach to numerically calculate the angular momentum profile at a later phase after the collapse has started. In Appendix \ref{sec:j_analytical}, we present the spatial profiles of $\Sigma(x)$ and $j(x)$ at $t=0$ for the tapered isothermal density profile of Equation (\ref{eq:rho_mod}), which are analytically tractable for this special case.
In the inner region where $M_*(t) > M(x)$, $M_{\rm tot}(x)$ will be nearly constant and approximately equal to $M_*(t)$. 
In Figure \ref{fig:Rout5_2D_radial_jMSigma}b the blue curves shows the specific angular momentum $j(x)$ as a function of $x/r_c$ at a time $t=0.35 \ {\rm Myr}$ after the collapse has proceeded and the central star is formed. We notice that within the inner region, $j(x)$ is almost flat up to $x/r_c \approx 0.15$ and then outside this region $j(x)$ increases linearly and comes to a saturation at the outer edge. 
This kind of $j(x)$ profile reasonably corresponds to the evolution starting
from the prestellar collapse (in the outer region) to a radial expansion wave regime (in the inner region) with $\Omega \propto r^{-2}$ \citep[see][]{dapp12}.
For a typical $r_c=0.037 \ {\rm pc}$, the break occurs at $x \approx 1021 \ {\rm AU} \approx 0.005 \ {\rm pc}$. The range of $j(x) \simeq 10^{20} - 10^{21} \ {\rm cm}^2 \ {\rm s}^{-1}$ and the shape of the profile is generally consistent with observations (e.g. fig. 6 of \cite{Ohashi1997}, fig. 13 of \cite{Yen_etal2011}, fig. 11 of \cite{KuronoEtal2013}, fig. 9 of \cite{Yenetal2017}, and fig. 17(b) of \cite{Gaudietal2020}).


\begin {figure}
\centering
\begin{tikzpicture}[node distance=1.35cm]
\node (s1) [startstop, text width =4 cm] {Calculate $\dot{M}_{\rm infall}$ from envelope};
\node (s3) [startstop, below of=s1] {Calculate $\dot{M}_{\rm ds}$ to estimate mass transportation rate from disc to the star};
\node (s4) [startstop, below of=s3, text width = 7.5cm] {
Total accretion rate on to star \\
$\dot{M}_*$= 10\% of $\dot{M}_{\rm infall}(t)$ +  90\% of the {\rm accretion via}\ $\dot{M}_{\rm ds}(t)$};
\node (s5) [startstop, below of=s4, text width=8.7cm] {
Update $\mdisc$ with 90\% mass from $\dot{M}_{\rm infall}$ -- mass transferred via $\dot{M}_{\rm ds}$,\\
Update $M_*$ with 10\% mass from $\dot{M}_{\rm infall}$ + 0.9  $\times$ mass transferred via $\dot{M}_{\rm ds}$};
\node (s6) [decision, below of=s5, yshift=-2cm, text width =3.2cm] {If $\mdisc/M_* \ > 0.33$ \\ (burst criterion)};
\node (s7) [startstop, below of=s6, text width=7.5cm, yshift=-2.3cm] {Episodic burst and mass transfer from disc to star through clump infall:
$\mdisc \rightarrow \mdisc - M_{\rm burst}$;    $M_* \rightarrow M_* + 0.5 \times M_{\rm burst}$};
\node (s8) [startstop, below of=s7, text width = 5.5cm] {We update $M_{0d}$, $t_{0d}$ after every burst, \\ calculate new $C_1$ and update $\dot{M}_{\rm ds}$; \\ (see steps (iv) to (viii) of Section 3.1)};
\node (s9) [startstop, below of=s8, text width = 6cm] {Total accretion rate on to star \\
$\dot{M}_*$ = 10\% of $\dot{M}_{\rm infall}(t)$ + 0.5 $\times$ $M_{\rm burst}/ \Delta t_{\rm burst}$};
\node (s10) [startstop, below of=s9, text width=6cm] {Proceed to the next time step and iterate from top};

\draw [arrow] (s1) -- (s3);
\draw [arrow] (s3) -- (s4);
\draw [arrow] (s4) -- (s5);
\draw [arrow] (s5) -- (s6);
\draw [arrow] (s6) -- node[anchor=east] {True} (s7);

\draw [black, thick] (s6) -- node[above] {False} (4.2,-7.4) ;
\draw [black, thick] (4.2,-7.4) -- (4.2,-15);
\draw [arrow] (4.2,-15) -- (s10);

\draw [arrow] (s7) -- (s8);
\draw [arrow] (s8) -- (s9);
\draw [arrow] (s9) -- (s10);

\draw [black, thick] (s10) -- (-4.6,-15.1) ;
\draw [black, thick] (-4.6, -15.1) -- (-4.6,0);
\draw [arrow] (-4.6,0.0) -- (s1);

\end{tikzpicture}
\caption{Flowchart of our formalism for the episodic accretion model. 
}
\label{fig:flowchart}
\end{figure}
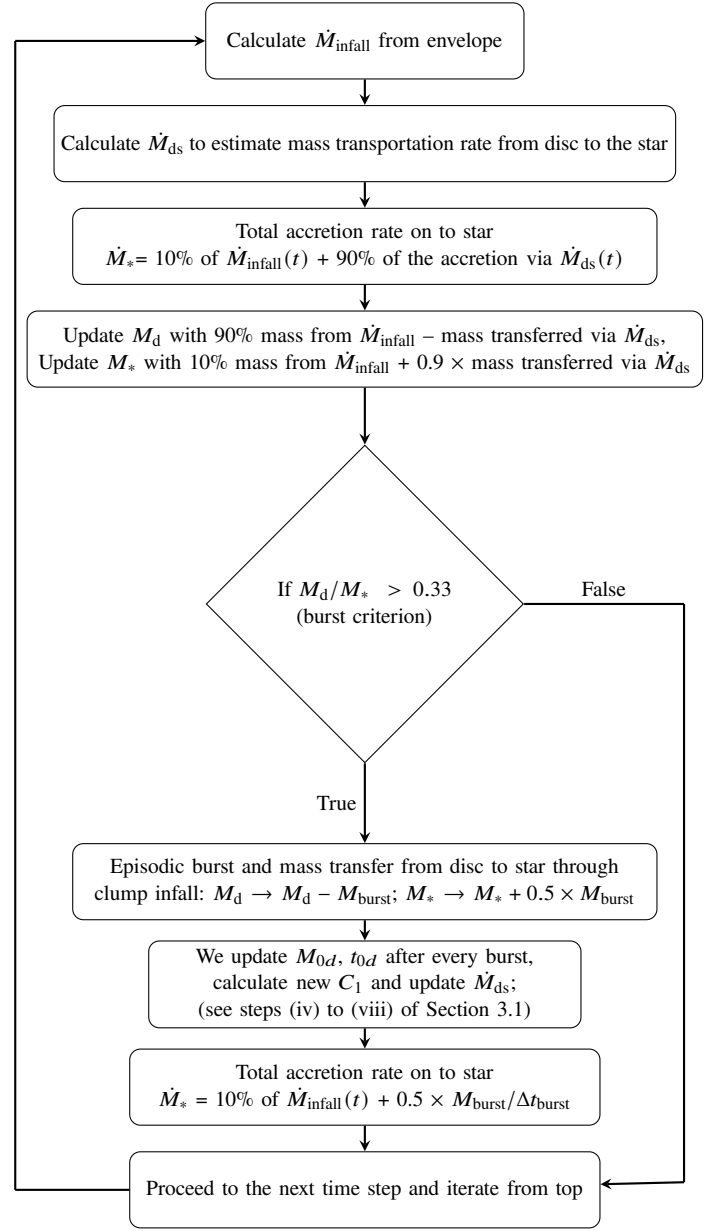




\subsection{Formalism of stellar accretion} \label{sec:sa_pres}
\begin{figure}
\includegraphics[width=0.95\linewidth]{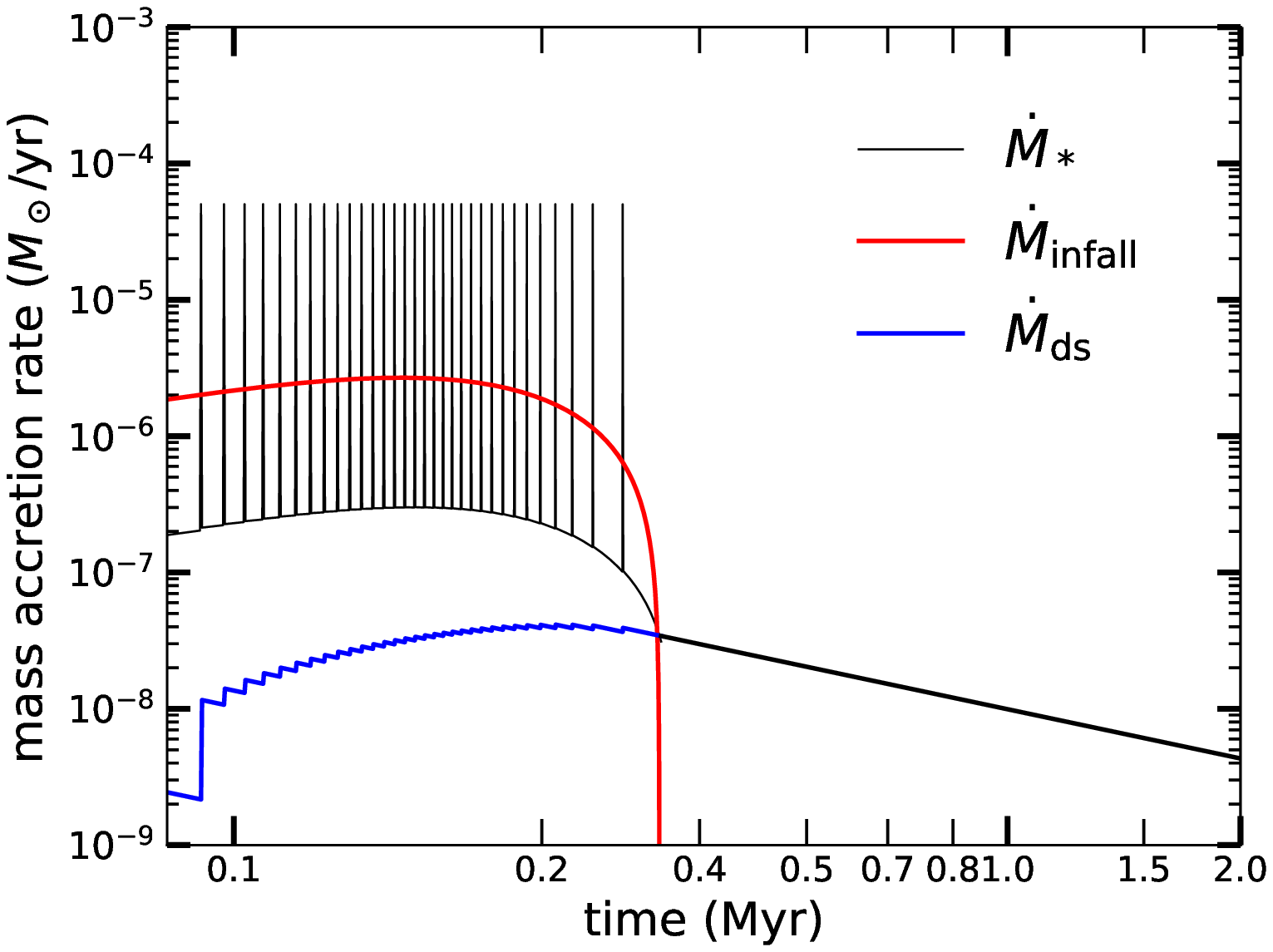}{(a)}
\includegraphics[width=0.95\linewidth]{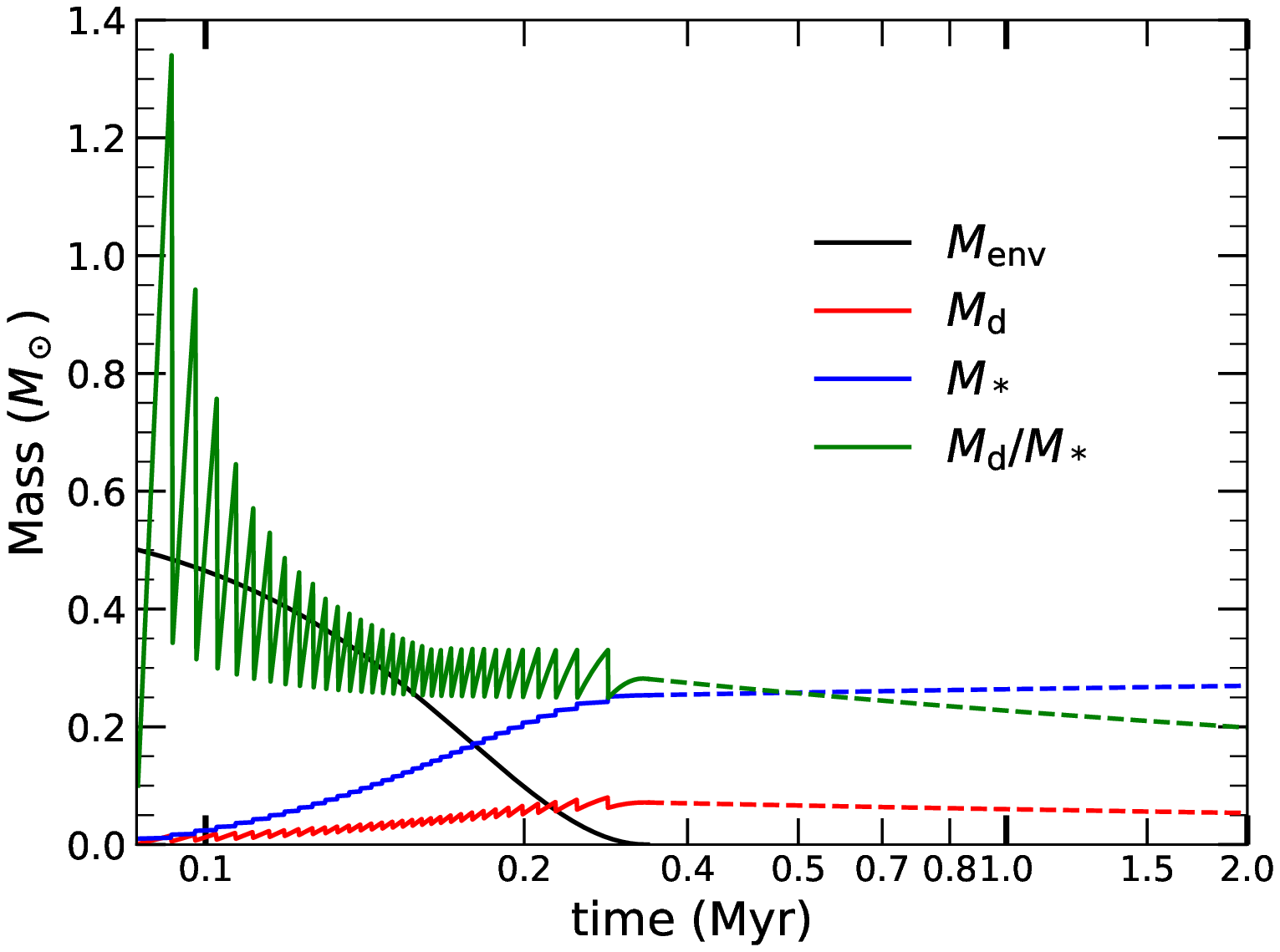}{(b)}
\includegraphics[width=0.95\linewidth]{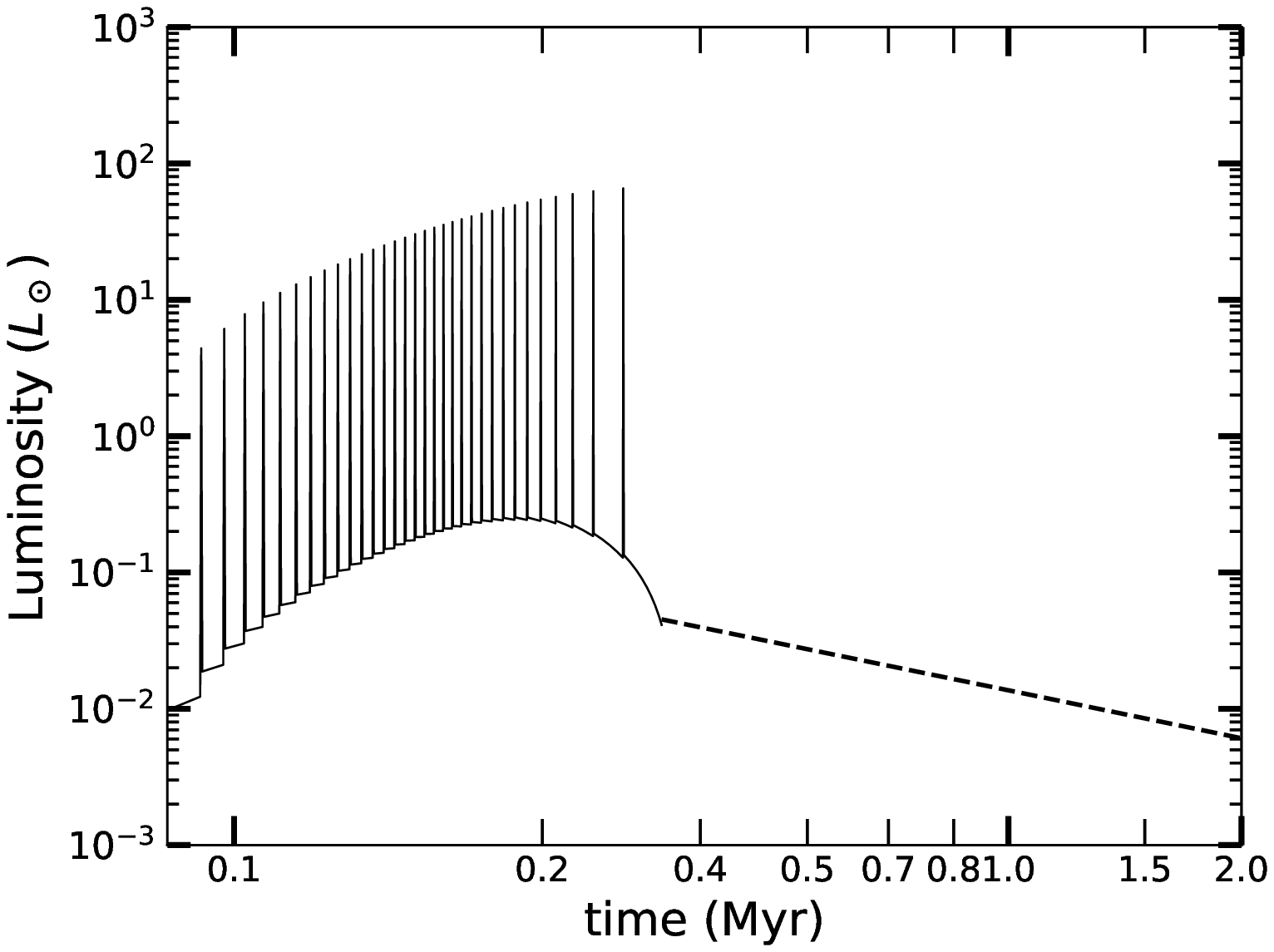}{(c)}
\caption{\textsc{model1}: (a) temporal evolution of the mass accretion rate (the total number of bursts is 32), 
(b) distribution of masses in the envelope-disc-star system, (c) temporal evolution of the accretion luminosity distribution.  
In (b) and (c), the solid line presents the joint evolution from envelope accretion together with disc accretion and the dashed line shows the evolution due to disc accretion only.}
\label{fig:model1}
\end{figure}

\begin{figure}
\includegraphics[width=0.95\linewidth]{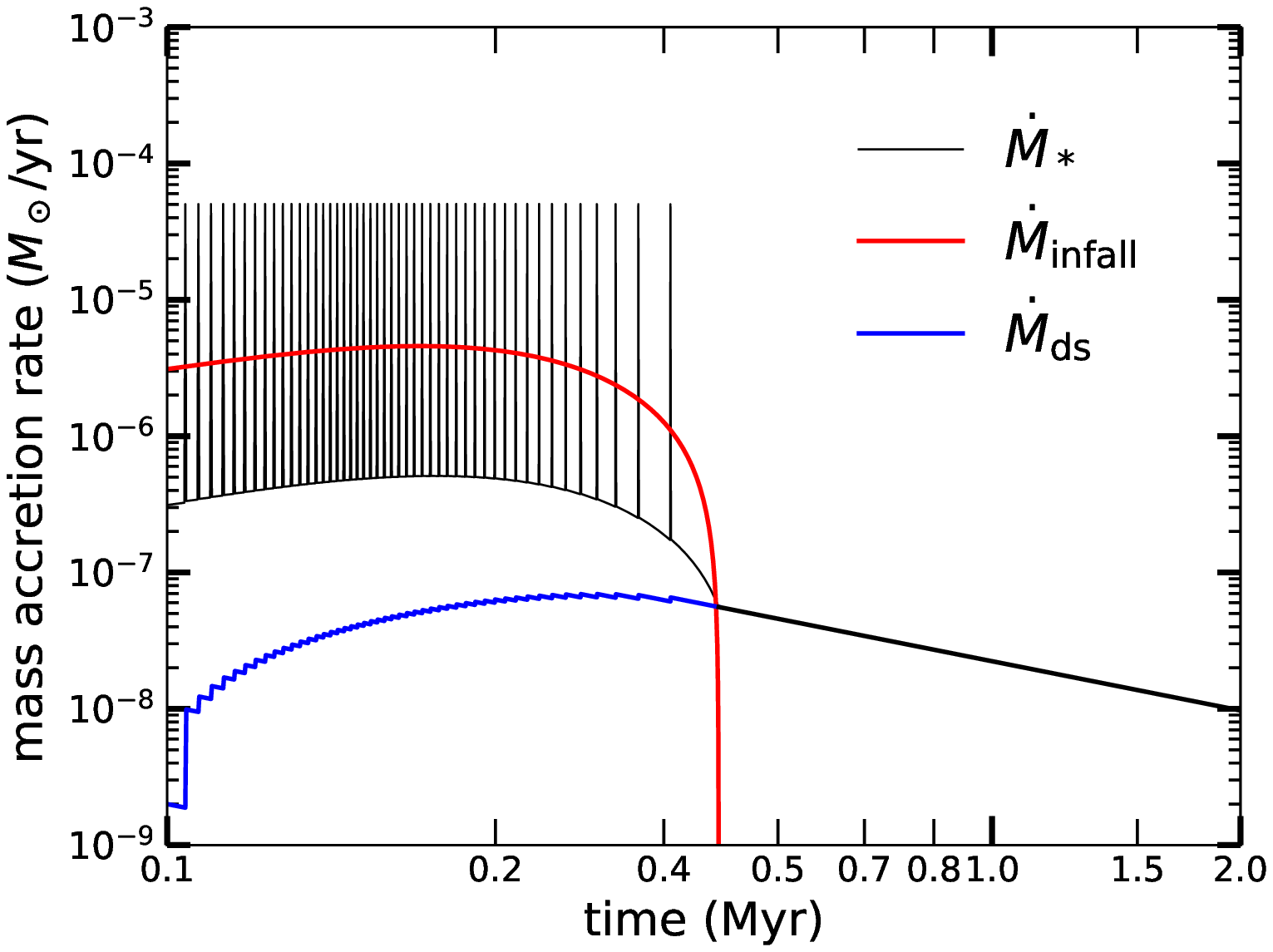}{(a)}
\includegraphics[width=0.95\linewidth]{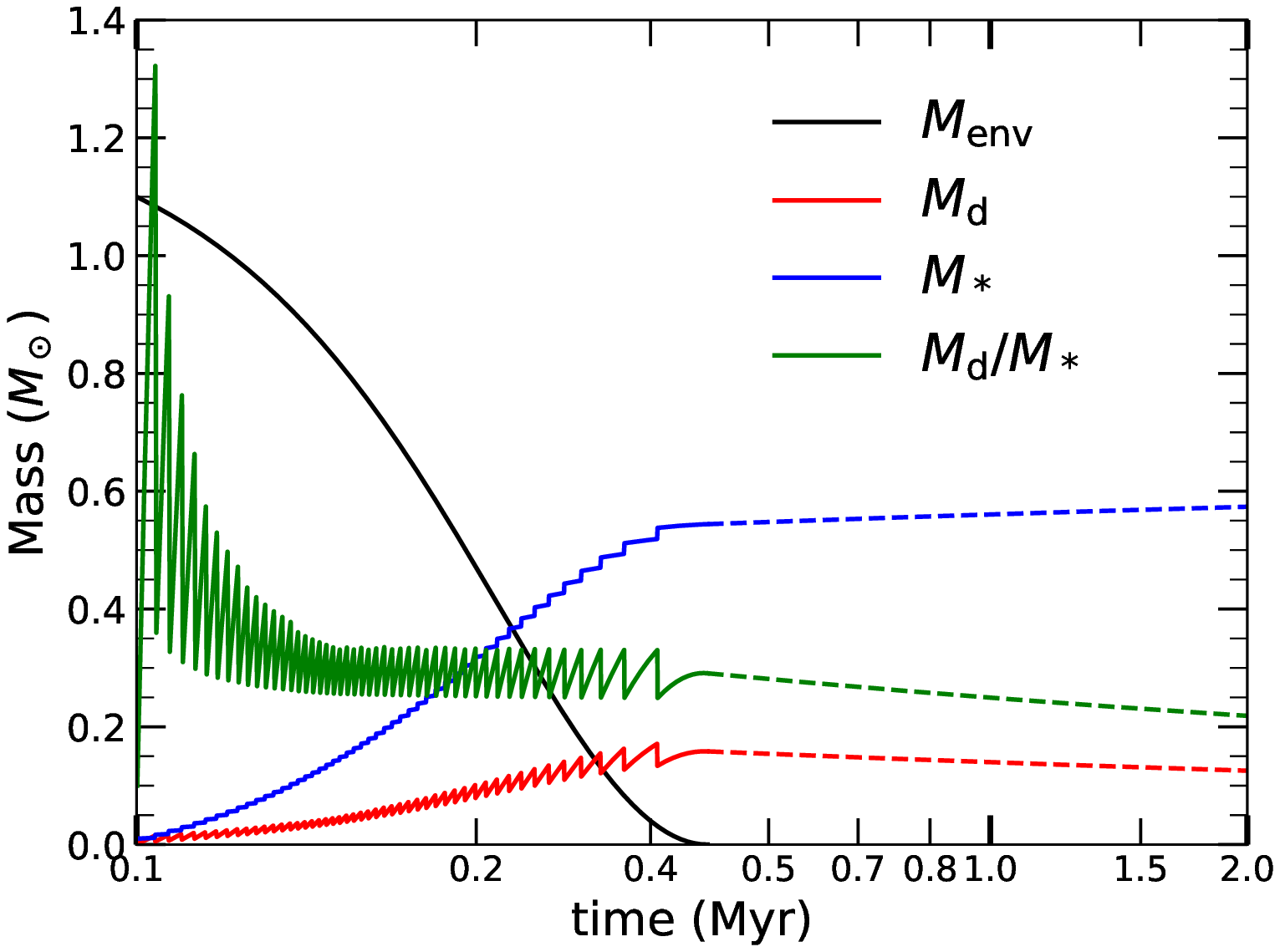}{(b)}
\includegraphics[width=0.95\linewidth]{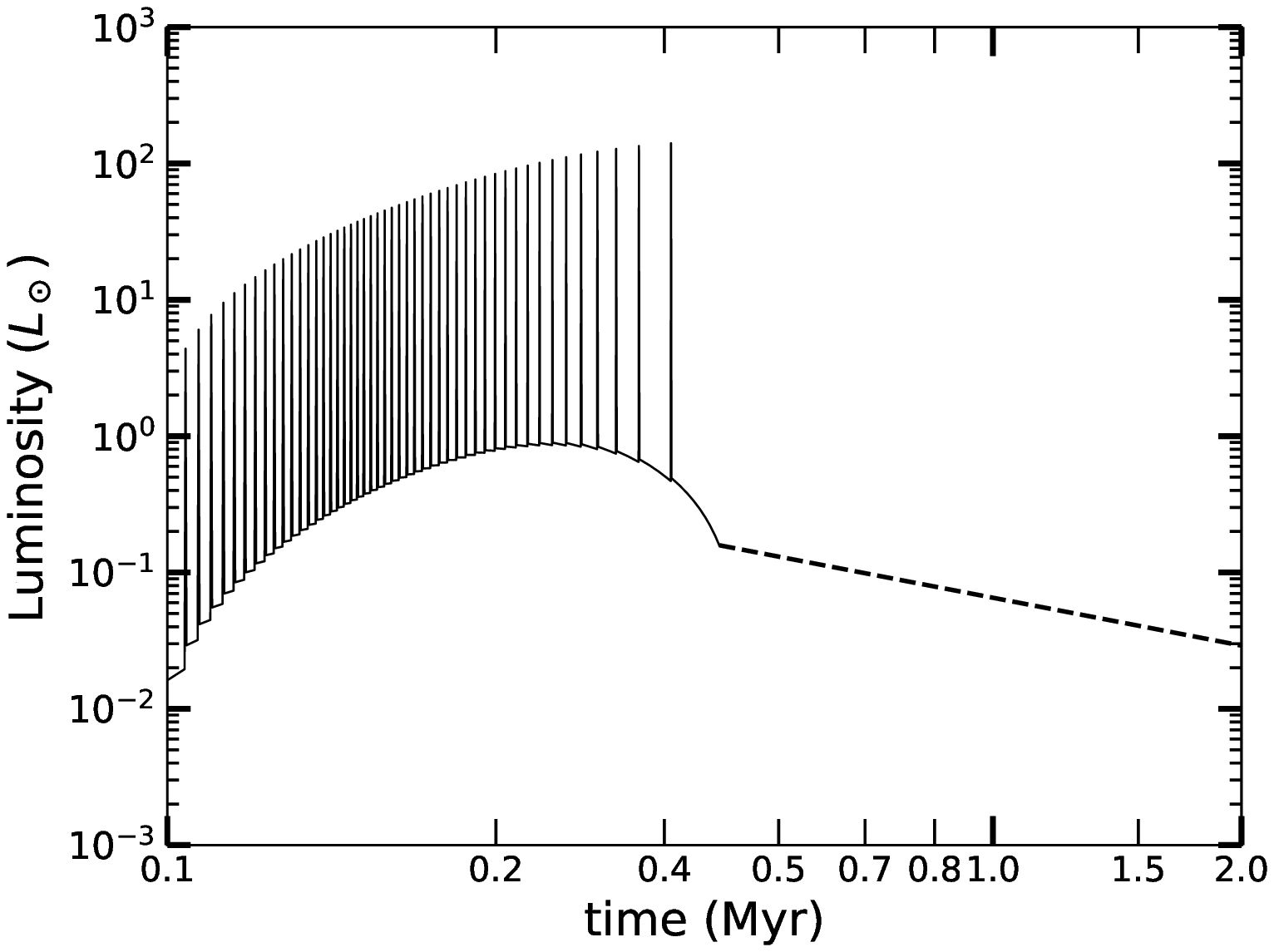}{(c)}
\caption{\textsc{model2}: (a) temporal evolution of the mass accretion rate (the
total number of bursts is 48), 
(b) distribution of masses in the envelope-disc-star system, (c) temporal evolution of the accretion luminosity distribution. 
In (b) and (c), the solid line presents the joint evolution from envelope accretion together with disc accretion and the dashed line shows the evolution due to disc accretion only.}
\label{fig:model2}
\end{figure}

\begin{figure}
\includegraphics[width=0.95\linewidth]{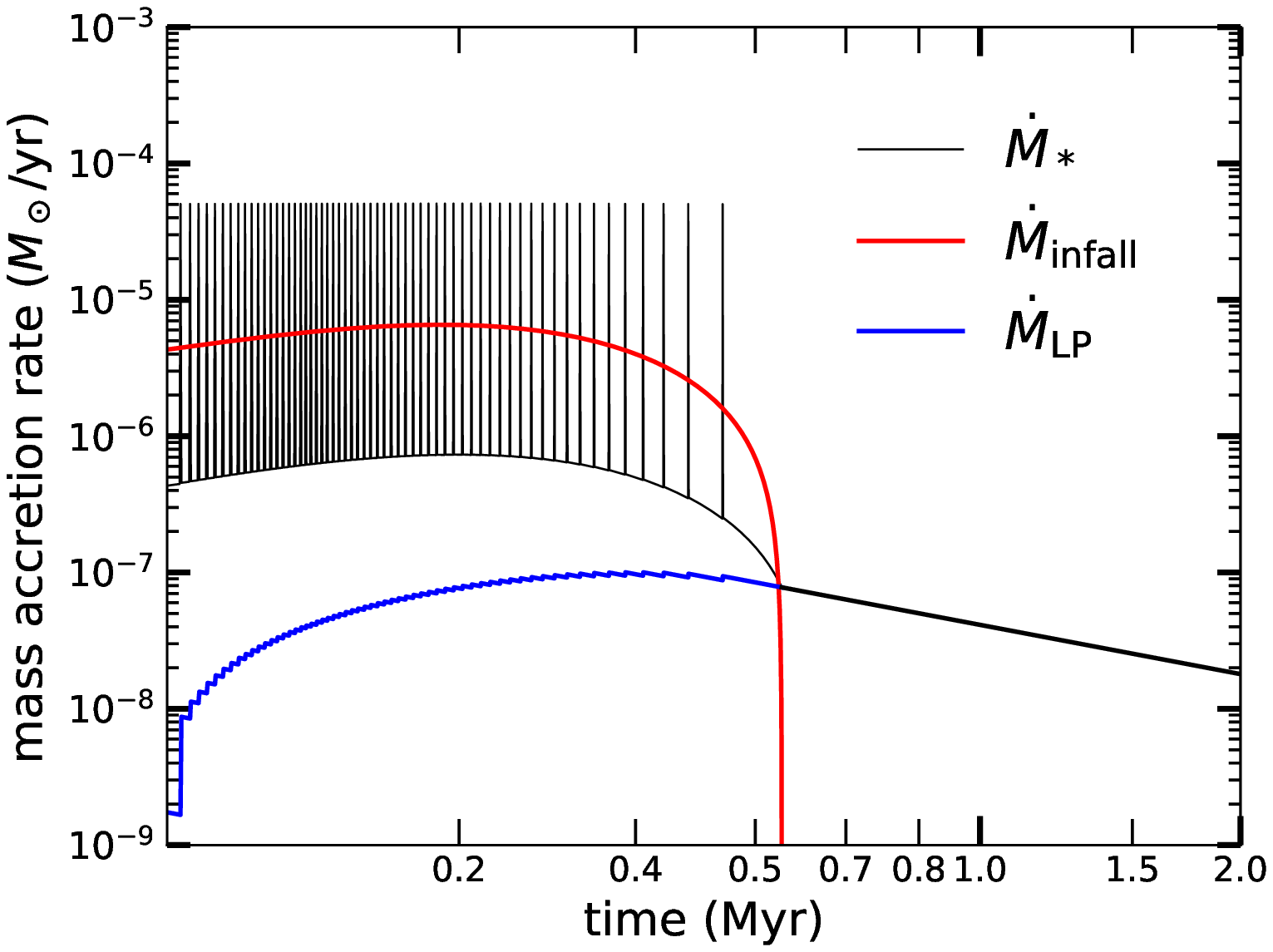}{(a)}
\includegraphics[width=0.95\linewidth]{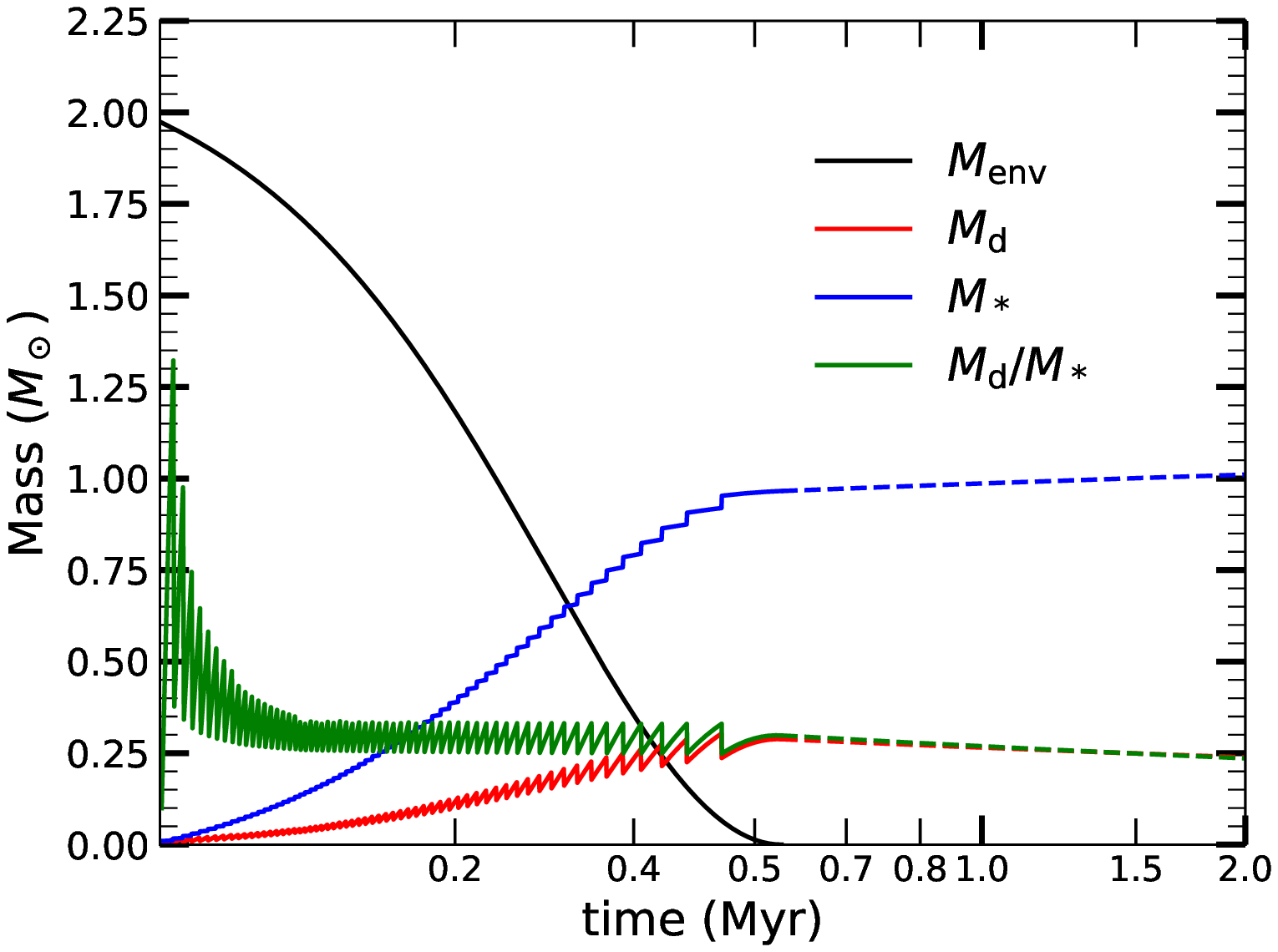}{(b)}
\includegraphics[width=0.95\linewidth]{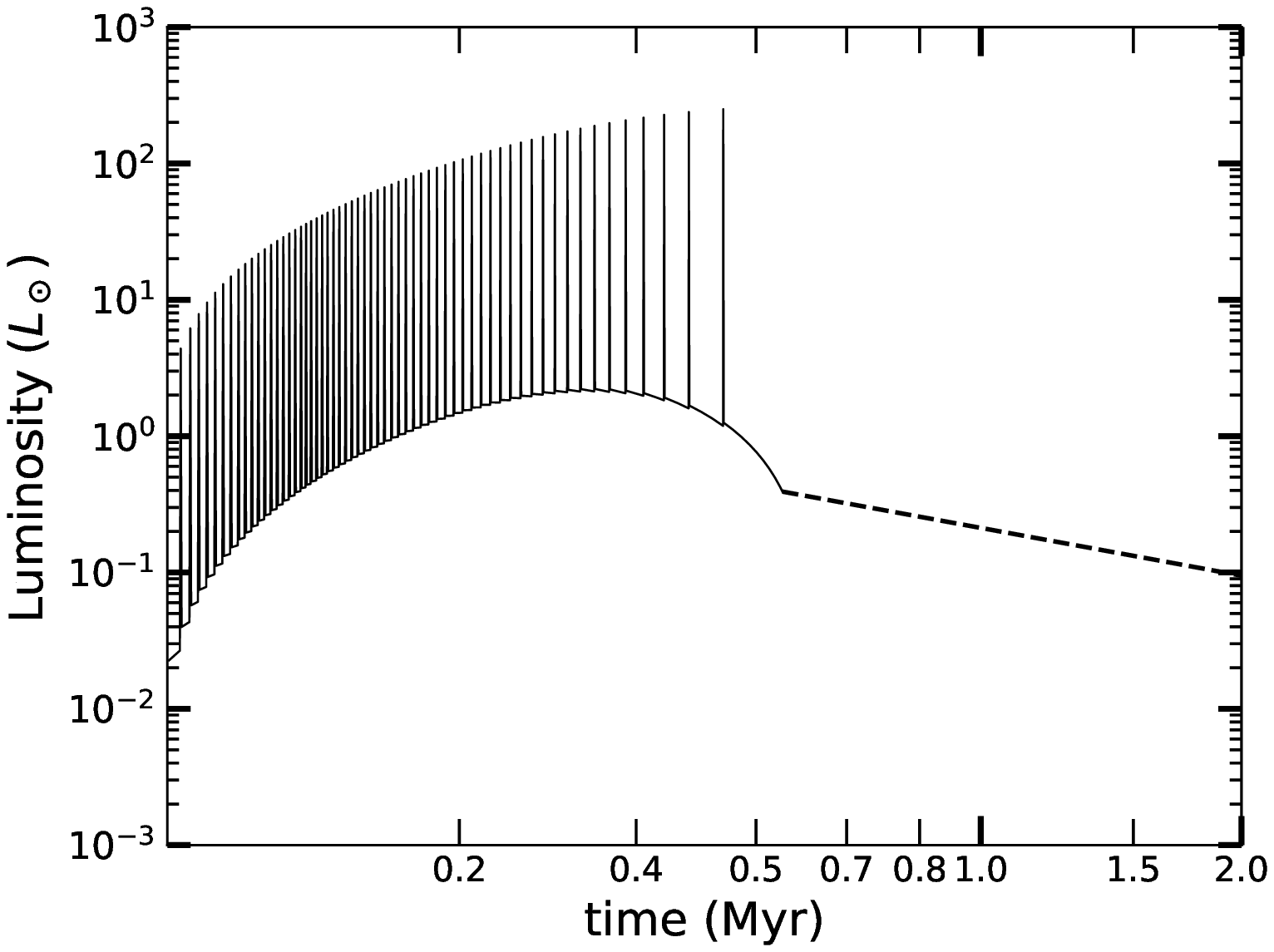}{(c)}
\caption{\textsc{model3}: (a) temporal evolution of the mass accretion rate (the
total number of bursts is 60), 
(b) distribution of masses in the envelope-disc-star system, (c) temporal evolution of the accretion luminosity distribution.  
In (b) and (c), the solid line presents the joint evolution from envelope accretion together with disc accretion and the dashed line shows the evolution due to disc accretion only.}
\label{fig:model3}
\end{figure}

In this section we describe (see Figure \ref{fig:flowchart}) how we obtain the mass accretion rate on to the star, $\dot{M}_*$, with episodic bursts as shown by the black line in Figures \ref{fig:model1}a, \ref{fig:model2}a, \ref{fig:model3}a. 
Here $\dot{M}_*$ is obtained from the joint result of the infall rate from the envelope and mass transportation rate from the disc to the star. 
We obtain the temporal evolution of the mass accretion rate at a radial distance $r=2r_c$ from envelope accretion for the models with different $R_{\rm out}$. 
We perform the following steps that demonstrate our semi-analytic formalism: \\
(i)  The initial masses of the disc and protostar are set to be $0.001\, \Msun$ and $0.01\, \Msun$, respectively. We consider $90\%$ of the total accretion from the envelope goes to the disc and the rest goes directly to the protostar (see Equation (\ref{eq:dotmstar})). At each time, the disc and star mass can be calculated by integrating the respective mass accretion rates from the envelope to the disc and the star, respectively. \\
(ii) Thereafter the mass transportation is calculated from disc to the star by integrating Equation (\ref{eq:nondim_mdotLP}). Note that, before the occurrence of the burst, $t_{0d}$ is considered as $t
_0$. We calculate $C_1$ accordingly as shown in Equation (\ref{eq:C1}). 
The amount of mass that the disc loses is added to the star. We then update the disc and the star masses. \\ 
(iii) If the disc-to-star-mass ratio $\mdisc/\mstar$ does not go above the threshold, then the disc is gravitationally stable and we calculate the $\dot{M}_*$ using only the contributions of the mass accretion rate from disc to star and from envelope to the star. \\
(iv) However, if the disc-to-star-mass ratio $\mdisc/\mstar$ goes above $0.33$ (and the disc mass is at least $0.01 \, \Msun$), the disc becomes gravitationally unstable. The disc transfers the matter to the star (which gives rise to an accretion burst) so that the $\mdisc/\mstar$ ratio goes below the threshold and the disc becomes gravitationally stable.  \\
(v) The amount of mass depleted with each burst is called the burst mass and we require that at least $0.01\, \Msun$ should be associated with a burst. During an accretion burst, the burst mass goes from the unstable disc toward the central protostar, adding to its mass. Outflows are taken into account such that  during the envelope accretion phase $50 \%$ of the mass accreted during a single burst episode goes to the outflows, while $10\%$ of the mass that is accreted via disc-to-star accretion goes to outflows. 
The molecular outflows are ubiquitous among protostars. The outflow energetics reflects a correlation to the mass infall/accretion rate  \citep[][]{BontempsEtal1996}.
Hence, during the disc accretion phase we 
update the stellar mass by taking $90\%$ contribution of the disc-to-star mass accretion rate $\dot{M}_{\rm ds}$ as mentioned in the third and fourth step of the flowchart (see Figure \ref{fig:flowchart}). 
Whereas, during episodic bursts $50\%$ of the burst mass gets finally accreted onto star as shown in the second last step of the flowchart.\\
(vi) After the burst, $C_1$ is calculated using the updated disc mass and $t_{0d}$. We update $t_{0d}$ to correspond to the most recent burst time and the disc mass is also updated accordingly as mentioned in point (iii).\\
(vii) Once a burst occurs, we calculate the burst mass as shown in Equation (\ref{eq:ds_ratio_rpd}) and the time duration as shown in Equation (\ref{eq:deltat_burst}). We calculate the mass accretion rate due to the burst by dividing the burst mass by the time duration of the burst. Then we add this to the pre-calculated mass accretion rate (from the disc and envelope) on to the star to calculate the final $\dot{M}_*$ at that time. \\
(viii) In between the bursts, the baseline of $\dot{M}_*$ is determined by the net mass accretion rate from the disc to the star 
and a little portion goes directly from the envelope to the star (see Equation (\ref{eq:mdot})). \\
(ix) We iterate the above steps as described in the points (i)-(viii) and also in Figure \ref{fig:flowchart} until the desired end of the evolution. 

The envelope accretion and disc accretion continue to coexist until approximately $99 \%$ of the envelope has been consumed by the protostar and protostellar disc. After that, the evolution of the mass accretion is set only by the disc accretion, which has a steady form $\dot{M}_{\rm ds} \propto t^{-6/5}$. Hence, when the envelope accretion ends, from then onwards $\dot{M}_*$ (black line) identically overlaps with $\dot{M}_{\rm ds}$ (blue line), as shown in Figures \ref{fig:model1}a, \ref{fig:model2}a, \ref{fig:model3}a. 
The accretion bursts do not occur during this phase.
Due to the lack of accretion from the envelope, the disc can not become sufficiently massive for the GI that leads to outbursts.



\subsection{Temporal evolution of the mass accretion rate} \label{sec:mdot_fig}
Figure \ref{fig:model1}a, \ref{fig:model2}a, \ref{fig:model3}a present the temporal evolution of the mass accretion rate for the models with different core masses as described in the Table \ref{tab:model}. 
In each of these figures, the red line shows the mass accretion rate from the spherical envelope accretion, which is similar to the curves shown in Figure \ref{fig:mdot_Rout_2D}. Moreover, this curve resembles the infall rate from envelope as calculated in the simulations. 
The black line shows the mass accretion rate on to the central star from the disc and directly from the envelope. The distribution of mass accretion rate from the envelope separately to the disc and to star is mentioned in Section \ref{sec:sa_pres}. The spikes represent the episodic accretion bursts. 
The amplitude of the mass accretion rate at each burst is calculated as the burst mass divided by its time duration (Equation (\ref{eq:deltat_burst})). 
The blue line shows the mass accretion rate from disc to the star via the LP formula (see Equation (\ref{eq:nondim_mdotLP})). There is a step-like increment at the occurrence of each burst, which corresponds to an updated intercept $C_1$ (see Equation (\ref{eq:C1})). The factor $C_1$ has an implicit time dependence through the updated disc and star mass at the time instance of a burst. 
The bursts occur due to the onset of gravitational instability within the disc. The disc is transporting material to the star via the LP formula. Simultaneously, the disc is gaining matter from the envelope infall at a different rate. Due to the mismatch between the infall rate and transport rate, the disc 
becomes sufficiently massive that the disc-to star-mass ratio exceeds the threshold for gravitational Toomre-$Q$ instability. Such a gravitationally unstable disc gets fragmented into spiral arms and clumps. The spiral arms are moving outward (gaining angular momentum) and the clumps are driven inward through the gravitational torques (losing angular momentum) and fall on to the central protostar. This leads to vigorous episodic accretion bursts that produce luminosity outbursts seen in observations.

Because of the finite mass reservoir, the envelope gets depleted at an earlier time for the lower mass cores.
At a later time, during the phase with only disc accretion, there is no burst, which delineates the simple power law profile where $\dot{M}_{\rm ds} \propto t^{-6/5}$ with a constant intercept.
We notice the total number of bursts is increasing with increasing core mass. We see the least number of bursts for \textsc{model1}. Because for the evolution of a small core of $0.5\, \Msun$, the disc does not get much time to continue to grow and spherical envelope accretion ceases early. 
Additionally, at the very initial time, the baseline of the $\mdot$ increases to $\sim 2 \times 10^{-7} \Msunperyr$, 
$\sim 3 \times 10^{-7} \Msunperyr$, $\sim 4 \times 10^{-7} \Msunperyr$ for \textsc{model1}, \textsc{model2}, and \textsc{model3}, respectively. 
We do not get any burst during the disc accretion phase, which starts typically after $t \approx 0.35 \ \Myr,$ $0.45 \ \Myr$, and $0.55 \ \Myr$ for \textsc{model1}, \textsc{model2}, and \textsc{model3}, respectively. 
Overall, we find that the evolution starts early for a low core mass and it starts at a later time for an increased core mass.


\subsection{Mass estimation} \label{sec:mass_fig}
Figures \ref{fig:model1}b, \ref{fig:model2}b, \ref{fig:model3}b present the masses contained in the envelope (black line), protostellar disc (red line), and the central protostar (blue line), for the models with different core masses as described in Table \ref{tab:model}. 
The mass in the envelope gradually falls over the time as it is being continuously absorbed by the protostellar disc and star.  
In the earlier stage, the spherical envelope accretion (the segments shown by solid lines) takes over the disc accretion (as shown by the corresponding dashed line) until $99 \%$ of the envelope mass has been depleted. 
We find that each sharp increase in the stellar mass is correlated with a corresponding sharp decrease in the disc mass. These sharp step-like increments/decrements happen because of the addition/subtraction of the finite burst mass to/from the star/disc, respectively, during every infall of a clump from the gravitationally unstable disc to the star. 
The accretion bursts continue to occur until the envelope mass has reduced by $90\%-95 \%$ and the mass accretion rate drops to $\sim 10^{-7} \Msunperyr$. 


Our study shows that at the end of the evolution the final stellar mass acquired is about $50\%$ of the initial parent envelope mass, which is approximately consistent with some observational estimates \citep[e.g.][]{alves2007}. 
Outflows also constrain the final stellar mass as found from the observations and theoretical models. 
In our model, on average $50-60 \%$ of the final stellar mass comes from the accretion bursts. 
However, the observations have been used to estimate that $10\%-35 \%$ of the total stellar mass is accumulated during the bursts. This is obtained for low mass star formation \citep{DunhamVorobyov2012} and also for high mass stars \citep{Caratti_etal2017, Meyer_etal2017,Magakian_etal2019}. 
In our study, episodic accretion plays a dominant role in the stellar mass growth as the disc accretion via the LP formula is not as vigorous as the episodic accretion. A more sophisticated model of mass and angular momentum transfer due to disc viscosity could help to refine the results. We keep this aside for future studies.
The green line shows the temporal evolution of the disc-to-star-mass ratio. At the early times, this ratio exceeds unity. Soon after that, the disc-to-star-mass ratio goes below unity, as found in observations. We note that 
our model is applied to very early times consistent with the Class 0/I phase, when this ratio may be greater than in the later (and more frequently observed) Class II phase when the envelope mass has largely dissipated and bursts are less frequent.

During the episodic bursts, the clumps of $\sim 0.01 -0.08 \ \Msun$ migrate inward to the centre. 
As the mass associated with a burst increases so does the time duration of the burst (see Section \ref{sec:mass_fig}). 
It happens because of the choice of a simple linear formula for fitting time duration with the mass for each burst as seen in Equation (\ref{eq:deltat_burst}); a typical burst of $0.01 \ \Msun$ corresponds to a duration of $100 \ {\rm yr}$. 
For \textsc{model1} with core mass $0.5\, \Msun$, the burst masses are ranging from $0.01-0.017\, \Msun$, with a time duration of $110-170 \ {\rm yr}$. 
For \textsc{model2} with core mass $1.1\, \Msun$, the burst masses range from $0.01-0.037\, \Msun$, with a time duration of $132-377 \ {\rm yr}$. 
For \textsc{model3} with core mass $\simeq 2\, \Msun$, the burst masses range from $0.01-0.05\, \Msun$, with a time duration of $132-500 \ {\rm yr}$ up to a time $t= 0.3 \ \Myr$. 
At a later stage ($0.3\, \Myr < t < 0.46\, \Myr$), for this relatively high-mass model, there are a few bursts (around $4-6$), and the burst mass goes up to $\sim 0.05 -0.08\, \Msun$ with a time duration of $500-830 \ {\rm yr}$. 
This amount of mass and time duration for the physical episodic bursts may seem excessive, however it 
can be thought of as the accumulation of several clumps of $\sim 0.01-0.02\, \Msun$ occurring in rapid succession, each with $\sim 100 \ {\rm yr}$ duration. Such clustered (knotty) bursts occur due to the disintegration of large clumps that are tidally disrupted as they approach the star, as described by \citet{vor15} and used by \cite{EV_etal_knottyjets2018} to explain the multiple knots in the vigorous jets from Herbig-Haro objects. 
The time resolution of our current scheme is $\sim 4 \ {\rm kyr}$ and we note that we do not resolve smaller intervals in order to keep our computation simple. Our aim is to reproduce the big picture of the main characteristics of mass accretion as found from the very time-consuming numerical hydrodynamic simulations.
Finally, we notice that sums of all masses remain constant, which implies that there is a global conservation of mass in our semi-analytic model.

\subsection{Estimating luminosity and comparing with observations} \label{sec:lum_fig}
Figure \ref{fig:model1}c, \ref{fig:model2}c, \ref{fig:model3}c present the temporal evolution of the accretion luminosity for three different core masses, as described in the Table \ref{tab:model}. 
In our formalism, we calculate the accretion luminosity as
\begin{equation}
L_{\rm acc} = f_{\rm acc} \ \frac{GM_{*}\dot{M}_*}{R_*}    \ ,
\label{eq:Lacc}
\end{equation}
where $\dot{M}_*$ is mass accretion rate on to the central protostar during spherical envelope accretion, $M_{*}$ and $R_*$ are the stellar mass and radius, respectively. We assume that a factor $f_{\rm acc}$ of the total incoming gravitational potential energy is converted to radiation at a shock front where the infalling material meets the star. 
Here, we take $R_* = 3\, \Rsun$ and $f_{\rm acc} =0.5$. 
The luminosity evolution reveals a peaked curve since
the product $M_* \dot{M}_*$ reaches a peak at a certain time. 
At a very early time, $\dot{M}_*$ has a high value and $M_*$ is still low, whereas at a later time, $\dot{M}_*$ decreases and $M_*$ has increased.

Apart from the accretion luminosity, we account for the contribution from the photospheric luminosity. The intrinsic photospheric luminosity can be expressed as
\begin{equation}
    L_{\rm phot} = 4\pi R_{\rm eff}^2 \sigma_{\rm SB} T_{\rm eff}^4 \ ,
\label{eq:Lphot}   
\end{equation}
where $\sigma_{\rm SB}$ is the Stefan-Boltzmann constant, and $R_{\rm eff}$ and $T_{\rm eff}$ are the radius and effective temperature of the stellar photosphere, respectively. In general, $L_{\rm phot}$ is much less than $L_{\rm acc}$ during the bursts. In between the bursts it is comparable to $L_{\rm acc}$. 
For the completeness of our study, we add the contribution from the photospheric luminosity based on the survey of stellar masses \citep{YorkeBodenheimer1993I, YorkeBodenheimer1995II} during the evolution, as used in the numerical simulations of \cite{EV_etal_MHD2020}.
\begin{figure}
\includegraphics[width=\columnwidth]{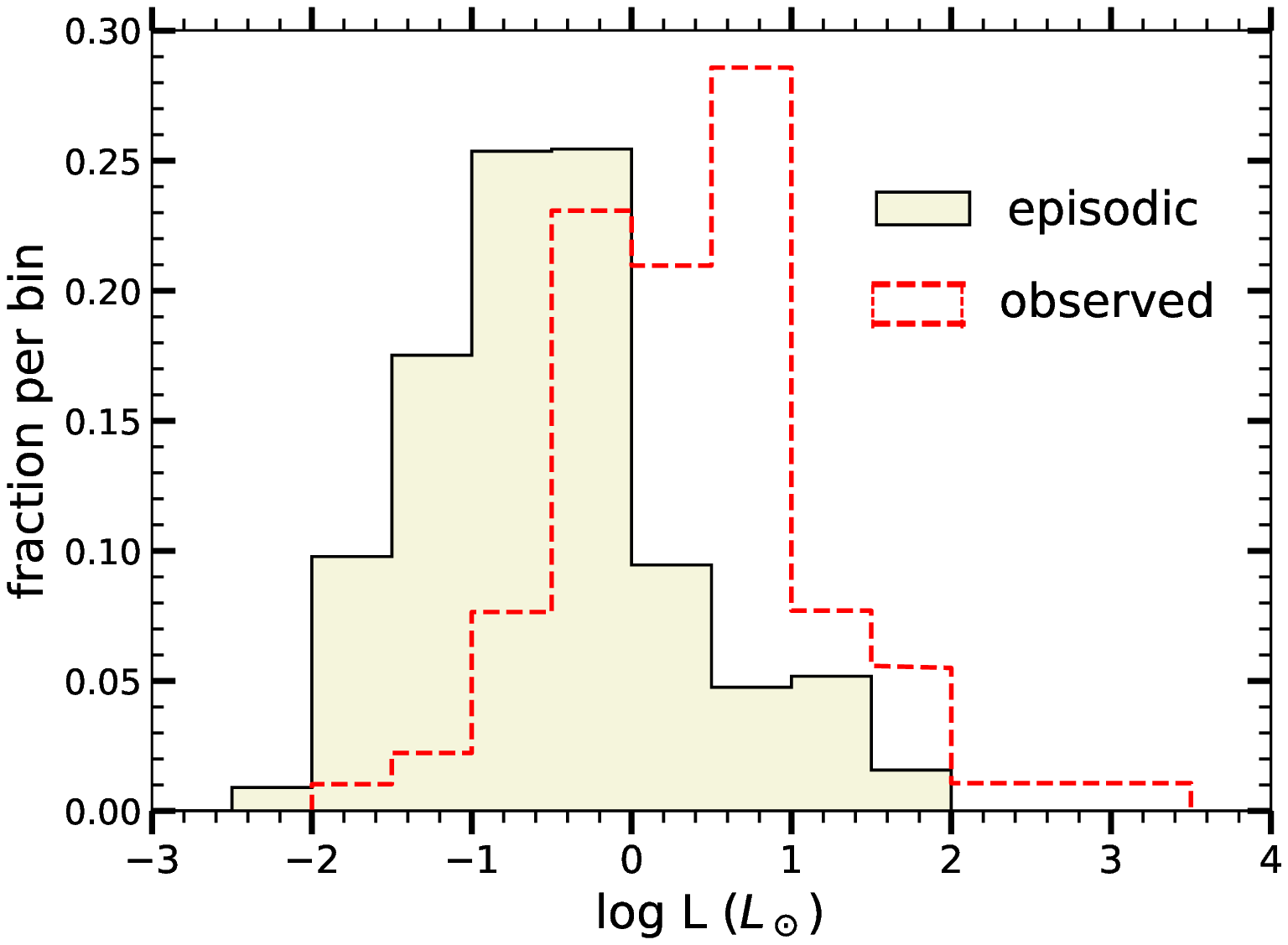}{(a)}
\includegraphics[width=\columnwidth]{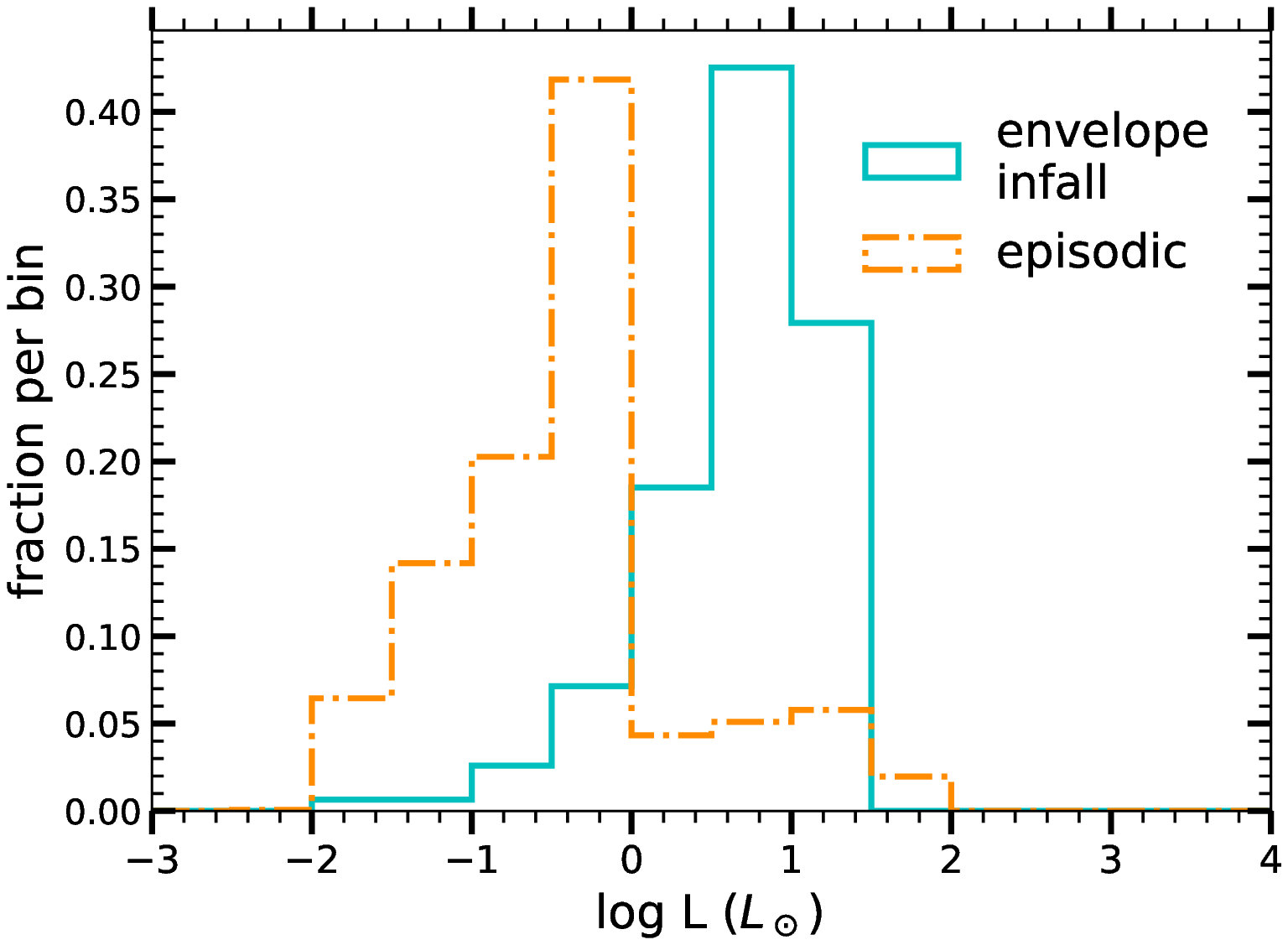}{(b)}
\includegraphics[width=\columnwidth]{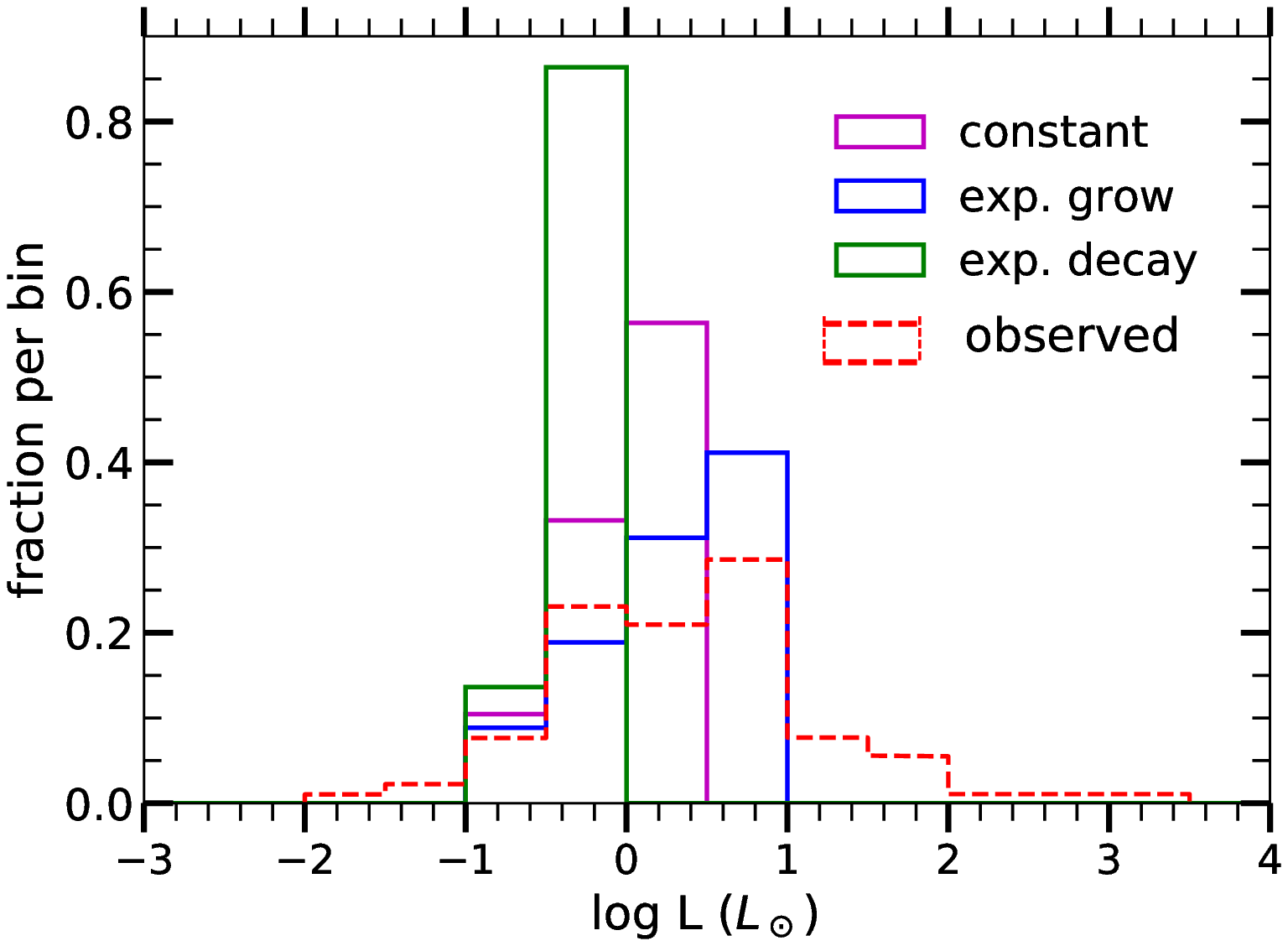}{(c)}

\caption{Top (a): Histograms of luminosities from our episodic model (the filled histogram), and the observations (red dashed line) as shown in the right panel of figure 2 of \citet{fischerEtal2017}. 
Middle (b): Histograms of luminosities from a single episodic model (\textsc{model2}) (orange dashed line) and its respective envelope infall model (cyan solid line).
Bottom (c): Histogram of luminosity for the analytic models with constant (magenta), exponentially growing (blue), exponentially decaying (green) mass accretion rate. The luminosity histogram from observations is shown in the red dashed line for comparison.}
\label{fig:Lhist_model3}
\end{figure}

We present a comparison of different luminosity distributions in Figure \ref{fig:Lhist_model3}a. 
The filled histogram shows the luminosity distributions for the Class 0 stellar objects within an evolutionary span of $\sim 0.1 \, \Myr$ based on all the models of different final stellar masses as shown in Table \ref{tab:model}. This time interval essentially corresponds to the time until the end of the embedded protostellar phase.
We consider this time range based on the mass remaining (about $50\%$) in the envelope for a typical Class 0 object. Once the object goes to the Class I phase, the envelope is diminished. 
To construct the luminosity histogram (the filled histogram in Figure \ref{fig:Lhist_model3}a, we consider models with a fairly large range of final stellar masses. 
We also take into account of the appropriate weights based on the initial mass function (IMF) for the different models categorized by the final stellar mass. These weights for the respective logarithmically spaced mass bins are calculated from the cumulative probability distribution presented in \cite{basu2015}. 
The red dashed histogram in Figure \ref{fig:Lhist_model3}a presents the luminosities of YSOs found from the \textit{Herschel} Orion Protostar survey as carried out by \citet[][see their fig. 2]{fischerEtal2017}. 
The observed histogram is drawn from
91 Class 0 objects.
Figure \ref{fig:Lhist_model3}b clarifies the difference in luminosity distribution due to episodic accretion in a single model. The histograms in cyan (with solid line) and orange (with dashed line) in Figure \ref{fig:Lhist_model3}b show the luminosity distributions obtained for \textsc{model2} calculated from the
spherical envelope infall (see the red curve of $\dot{M}_{\rm infall}$ in Figure \ref{fig:model2}a) and the calculated episodic accretion (see the black curve of $\dot{M}_{\rm infall}$ in Figure \ref{fig:model2}a), respectively. 
The histogram obtained from the spherical envelope infall contains many fewer low luminosity values, corresponding to the `luminosity problem' when the theoretical mass accretion rate $c_s^3/G$ is compared with the observed rate. 
In a realistic situation with a rotating core, the nonrotating spherical model value $c_s^3/G$ can be thought of as the envelope accretion on to the disc.

In the filled histogram as obtained from our episodic model and shown in Figure \ref{fig:Lhist_model3}a, the range of luminosities from 
$\sim 10\ \Lsun$ to $\sim 100 \ \Lsun$ correspond to the episodic bursts. 
On the other hand, luminosities around $10^{-2} \ \Lsun$ are associated with the early time evolution. 
However, 
such faint objects
are hard to detect observationally. 
We compare the observed and theoretically obtained histograms quantitatively by using the histogram 
intersection method.  This method calculates the similarity 
between two distinct histograms by adding the minimum fraction from each bin, with a maximum possible value of 100\% if the histograms are identical. A comparison in this case yields a reasonably good cumulative value of 55\%.
Sampling over many episodic models of different stellar masses and ages might provide a better match 
to the peak of the observed histogram and can be pursued in future work.

To investigate the effect of accretion bursts in the luminosity distributions, we also study the similar diagnostics with a constant mass accretion rate as well as time dependent but smoothly increasing and decreasing mass accretion rates. 
Figure \ref{fig:Lhist_model3}b shows the histogram of luminosity distributions for analytic models with constant, exponentially growing, and exponentially decaying mass accretion rate, respectively. 
The models are described by
\begin{equation}
  \dot{M}(t) =
    \begin{cases}
      \dot{M}_0 & \text{constant (a),}\\
      \dot{M}_0 e^{(t-t_a)/\tau} & \text{exponentially growing (b),}\\
      \dot{M}_0 e^{-(t-t_a)/\tau} & \text{exponentially decaying (c),}
     \end{cases}  
\label{eq:ref_model}     
\end{equation}
where $\dot{M}_0$ is the initial mass accretion rate $2 \times 10^{-6} \, \Msunperyr$, $\tau$ is the growth or decay constant for the evolution of mass accretion and is set to $0.1 \; \Myr$, $t_a$ is the initial time $0.1 \; \Myr$.
We evolve these models from $0.1 \ {\rm \Myr}$ to $0.2 \ \Myr$ with the initial $M_*=0.01\, \Msun$. 
We choose this value of $\tau$ so that the mass of the YSO at sampled time 0.2 Myr is $0.14\, \Msun$ for the decaying solution and $0.35\, \Msun$ for the growing solution. These values contain approximately the expected range of Class 0 object masses in low mass star-forming regions. Substantially different values of $\tau$ can lead to a range of protostar masses that are too small or too large. Of course, a more comprehensive study of growing and decaying mass accretion models remains to be done and is beyond the scope of our present work.
We find that the range of luminosity distributions for our models are 
$\sim 0.1-1 \ \Lsun$, $\sim 0.1-4 \ \Lsun$, $\sim 0.1-10 \ \Lsun$ for exponentially decaying, constant, exponentially growing mass accretion rates, respectively.
We notice from Figure \ref{fig:Lhist_model3}c that the entire range (i.e. from the low to high end) of luminosities for the above mentioned analytical models is less than that of the observations. 
The range of the observational luminosity distribution (the red dashed line in Figure \ref{fig:Lhist_model3}a) is larger by a few orders of magnitude than that predicted with the constant or smoothly decreasing and increasing mass accretion rates. 
Note that the luminosity histogram for the model with exponentially decaying mass accretion rate covers the shortest range of luminosities ($0.1\, \Lsun \lesssim L \lesssim 1\, \Lsun$), and is the least viable model. 
We see that the histogram of total luminosities obtained from our semi-analytic episodic accretion model provides a better fit to the observed histogram.
It implies that episodic accretion bursts are required, at least at this initial stage, in order to reach solar-type masses within a decent time interval of $\sim 0.1 \ {\rm Myr}$. 
Episodic accretion aids in accumulating $\sim 35 \%$ of the final stellar mass within a timeline of $\sim 0.1 \, \Myr$ (as seen in Figure \ref{fig:model3}b). 

For the models with exponentially decaying, constant, and exponentially growing mass accretion rates,
the accumulated masses within a standard protostar formation timeline of $0.1 \ \Myr$ are $\sim 0.14, 0.21, 0.35 \ \Msun$, respectively. 
It indicates that simple mass accretion models (without the accretion outbursts) are not sufficient to simultaneously explain the observed mean mass accretion rate, accumulated mass and luminosities of YSOs.
A stellar mass accretion model incorporating the vigorous mass accretion bursts is a viable way of reconciling the observed star formation time and mean luminosity. 


\section{Discussion}\label{sec:diss}
Our semi-analytic model captures the main physical insights of the evolution of episodic
mass accretion rate during star formation. 
Our prescription captures some basic characteristics such as the evolution of the stellar and disc masses, the episodic accretion bursts, the burst amplitudes and corresponding luminosities. These are obtained in a simple computational scheme that is consistent with the results obtained from detailed hydrodynamic simulations of episodic accretion 
\citep[][see also \cite{vor10,vor15}]{vorobyov07}.

The prestellar core is thought to be threaded by a magnetic field with a supercritical mass-to-flux ratio, so that the magnetic field is weaker than self-gravity and a magnetically-diluted gravitational collapse can happen \citep[e.g.][]{Basu97, basu04}. In this work we ignore the effect of this magnetic field and work in the limit of spherical envelope accretion. We also ignore any modification to the Toomre criterion in the disc due to the magnetic field \citep{DasBasu2021}. 
Our approach in this paper has some conceptual similarity to the work of \citet{Terebey_etal_1984}, who modeled the quasi-spherical collapse solution of a freely-falling singular isothermal sphere with a perturbational addition of rotation. They also joined the solution at small radii to a disc evolution model. We have studied pure spherical collapse and treated angular momentum as only a passively advected quantity with no dynamical back reaction. Our disc evolution model differs from that of \citet{Terebey_etal_1984} in that it features the influence of GI and the occurrence of accretion bursts in the early evolution, and at late times is set by a model of gravitational torques that follows a mass accretion rate time dependence $\propto t^{-6/5}$.  

We have derived a profile of specific angular momentum (Figure \ref{fig:Rout5_2D_radial_jMSigma}b) as found in the observations as measured from ${\rm C}^{18}{\rm O}\, (J =1-0)$ in the core and ${\rm H}^{13}{\rm CO}^{+}\, (J =1-0)$ in the envelope \citep[][]{Ohashi1997, Yen_etal2011, KuronoEtal2013}.
A break point in the profile of specific angular momentum may be related to the transition from an inner density profile $\rho \propto r^{-3/2}$ of dynamical free-fall collapse to an outer profile $\rho \propto r^{-2}$ characteristic of near-equilibrium conditions. This has been investigated by \citet{KuronoEtal2013} and has been seen in theoretical models \citep[e.g.][]{Terebey_etal_1984,dapp12}. 
In our models, we take the cloud radius $R_{\rm out}$ to be $5 r_c$, and the transition to a $r^{-2}$ power law in the density profile is not distinguishable since the density falls sharply near the outer edge. Extending the outer edge of cloud might help to obtain a more distinct power-law break in the density in addition to that in the specific angular momentum along the line-of-sight. 
Additionally, assigning a rotation profile $\Omega(r)$ to each mass shell during the runaway collapse and studying the evolution self-consistently might be a more realistic way to probing the break in the specific angular momentum during the protostellar collapse.


During the earliest evolution, protostars are embedded and heavily obscured by a dusty envelope. 
The high extinction inhibits a determination of protostellar evolution, lifetimes, and the accretion process until the envelope mass reduces to $\sim 50\%$ of the original. 
We find that during the very early stage, the ratio $\mdisc/M_*$ goes above $1$ during the early evolution of the protostar (see Figures \ref{fig:model1}b, \ref{fig:model2}b, \ref{fig:model3}b), as also found in some numerical simulations 
\citep[e.g.][]{MachidaBasu2019, xu2021}. 
We consider $10\%$ of envelope infall goes directly to the star, i.e. it occurs along polar regions and not through a disc. 
By increasing this fraction, the stellar mass growth will be relatively faster and the disc mass can not grow as substantially. The disc would become less gravitationally unstable and less likely to have episodes of accretion bursts. 
Our semi-analytic model yields the evolution history even for the early embedded phase, which is important to explain and predict the later phase of star formation.


Work by \cite{ArmitageEtal2001epi} and \cite{BaeEtal2013epi} has modeled the disc evolution in a semi-analytic manner using a constant or exponentially decreasing external mass infall rate on to the disc. The disc evolves due to viscous terms that model either GI or the magnetorotational instability (MRI) in different regions. These disc models distinguish between a GI-driven outer region and an inner region controlled by MRI. The GI-driven evolution leads to a pileup of matter in the inner region that is then delivered in episodic bursts to the central star through the activation of the MRI. In our work, we have focused on the large-scale picture of episodic accretion by building a model for the envelope accretion and the evolution of the major part of the disc that is driven by GI and gravitational torques. This assumes that the burst properties are set by the physics of the outer disc and we do not include any modulation by the inner disc physics including MRI. Inclusion of a separate model for the inner disc is beyond the scope of our current study but remains a target for the future. Our model is meant to provide a semi-analytic means to model the numerical simulations of e.g. \citet{vorobyov06, vor10, vor15}, who modeled disc evolution with envelope infall but had a central sink cell of size several AU that precluded a study of the inner disc physics. 

In our work, we have modeled the histogram of luminosity distribution during a limited time range ($\sim 0.1 \ {\rm Myr}$) of the Class 0 and early Class I phase. 
The choice of such an evolutionary period corresponds to the typical time until the end of the embedded Class 0 phase when about half of the envelope is accreted. 
Comparing the histogram of luminosity distribution during this time with the observed luminosity histogram for these phases
\citep{fischerEtal2017},
we find that the episodic accretion is needed in order to provide a reasonable match. 
In our model, we neglect the luminosity contribution from the disc emission as we expect it to be $3-4$ orders of magnitude lesser than either the photospheric or accretion luminosity.
As an example many simulations \citep[e.g.][]{vorobyov07,EV_etal_knottyjets2018, EV_etal_MHD2020} show an elevated temperature $T\gtrsim 40$ K at radius $r \lesssim 14$ au, and this region would dominate disc emission from larger radii and is also well within the beam size of infrared telescopes used to observe Class 0 objects in the nearest star-forming regions. The estimated contribution to bolometric luminosity $\sim \sigma_{\rm SB} T^4 \pi r^2$ using the above values is only $\sim 10^{-3}\,\Lsun$.
We also do not include an excess luminosity from external heating of the envelope by the interstellar radiation field. This is more likely to be significant for Class 0 objects since they have more envelope mass to heat, and could dominate at the very early stages with very low mass and luminosity central objects. However, we do not include this effect in our luminosity histograms due to the lack of precise estimates of its value. 

In a different study, \cite{OffnerMckee2011} used models of either constant or time-dependent (initially increasing and then tapering off) mass accretion rate and found that the time-dependent models could in some cases provide an adequate fit to the luminosity histogram. Their best fit required an ensemble of objects that had a longer age spread ($\sim 0.3$ Myr) than generally assumed and an accelerating star formation rate. These effects increase the number of observed young low mass and low luminosity objects. Their model also included a modest degree of episodic accretion. Our mass accretion model contains the initially near constant accretion rate along with bursts and the later tapered accretion rate, due to depletion of envelope accretion as well as the natural decline of disc accretion due to internal torques. A declining accretion rate is a feature of all simulation models that have a finite mass reservoir \citep{VB05mnras_a}, as well as being inferred from observational analysis of outflow activity \citep{BontempsEtal1996}. 
However, it is also the case that to create high-mass protostars, which are not in the observational samples of \cite{DunhamVorobyov2012} and \cite{fischerEtal2017}, a period of exceptionally high mass accretion rate seems required. This is because the total time to form stars of all masses within a stellar cluster seems to be nearly constant \citep{myersfuller1993}. Using numerical MHD simulations of star cluster formation, \citet{Wang2010} found that massive stars form in the central region of a cluster as a result of global gravitationally-driven flows that are temporally increasing. If this is correct, then one can add a period of temporally increasing accretion rate at later times to account for the relatively small fraction of massive stars. A series of works on the stellar initial mass function have assumed an exponentially increasing accretion rate together with equally likely stopping of accretion in each time interval that allows only a small fraction of protostars to reach the high accretion rate phase and become high mass stars \citep{myers2000, basuimf2004, myers2011, myers2014, basu2015, hoffmann2018, essex2020}. 
A variety of accretion scenarios, for example, either a long-term lower-amplitude variation or a combination of long-term lower-amplitude variation (i.e. secular variability) and short-term high amplitude accretion (i.e. stochastic variability or episodic accretion) can also be plausible at matching the observed luminosities of protostars \citep[see review by][]{PPVIIaccretionvariability}.  Future observations are essential to place tighter constraints on the luminosity evolution, and therefore elucidate the mass assembly history of protostars of all masses. 

In future work, we plan to generalize our model to perform a large parameter survey to understand/portray more systemically the physics of mass accretion during star formation. Inclusion of other transport mechanisms such as MRI together with the existing GI approach might give a more complete picture.
One can take a random sample of prestellar cores of different masses and study the long-term evolution ($\sim 2 \ {\rm Myr}$) of their protostellar discs and protostars. 
One can then compare the luminosity distribution for an ensemble of objects of different masses and ages with that of observations.  

%
\section{Summary}\label{sec:sum} 
We have presented a semi-analytic model for the temporal evolution of episodic disc-to-star mass accretion rate during star formation.  
Our formalism can explain the basic features of the hydrodynamic simulations of episodic accretion \citep[][]{VB05mnras_a, vorobyov06, vorobyov07, EV_etal_MHD2020} as discussed in Section \ref{sec:results}.
In doing so, it provides an intuitive understanding of the detailed nonlinear physics of the simulations in terms of basic physical principles. The model captures the evolution of the masses of the disc, star, and envelope, as seen in Figure \ref{fig:model1}, \ref{fig:model2}, and \ref{fig:model3} \citep[as compared to fig 6a of][]{vorobyov06}. Our formalism combines the spherical envelope accretion with the episodic disc accretion for determining the mass accretion rate on to the star. The former is calculated from the collapse of the prestellar isothermal cloud core for a given density and velocity profile, which controls the infall of the material from the envelope to the disc and a very little portion (about $\sim 10\%$ of the infalling matter) directly on to the centre.
The latter governs the mass transport from the disc to the star, which follows a power law $\dot{M}_{\rm ds} \propto t^{-6/5}$.
We incorporate the episodic accretion bursts due to gravitational instability within the disc by tracking the disc-to-star mass ratio.
In our model, both the envelope and disc accretion are co-existing. The envelope accretion dominates the disc accretion in the earlier stages. However, in the later stage, when the envelope mass is depleted, episodic bursts cease to occur and the disc accretion becomes dominant. From then on, the disc accretion solely determines the mass accretion rate on to the central protostar.  

Our model includes of the effect of the episodic accretion bursts for determining the mass accretion rate and henceforth the accretion luminosity. 
We find that a constant or even an exponentially growing or decaying mass accretion rate is not sufficient to produce solar-mass type stars within a typical protostar formation time $\sim 0.1 \ {\rm Myr}$. The episodic bursts work through the migration of clumps of $\sim 0.01 \ \Msun$ from the disc to the centre and helps to accumulate enough mass at the centre within a desirable time frame, in contrast to a steady or exponentially increasing or decreasing accretion rate.
Our results (e.g. Figure \ref{fig:Lhist_model3}) indicate that the bursts are required to obtain a good fit to the observed distribution of bolometric luminosities of YSOs, as compared to a smoothly increasing or decreasing evolution of the mass accretion rate.  The episodic accretion seems necessary to explain the
long-standing  `luminosity problem'. 

\section*{Acknowledgements}
We thank the anonymous referee for the constructive comments that improved our manuscript.
We also thank Kundan Kadam for helpful discussions. This work was supported by a Discovery Grant from the Natural Sciences and Engineering Research Council of Canada.

\section*{Data Availability}
The data underlying this article will be shared on a reasonable request
to the corresponding author.



\bibliographystyle{mnras}
\bibliography{myref} 




\appendix
\section{Analytic solutions for mass and specific angular momentum profile} \label{sec:j_analytical}

\begin{figure}
\centering
\centering
\includegraphics[width=0.95\linewidth]{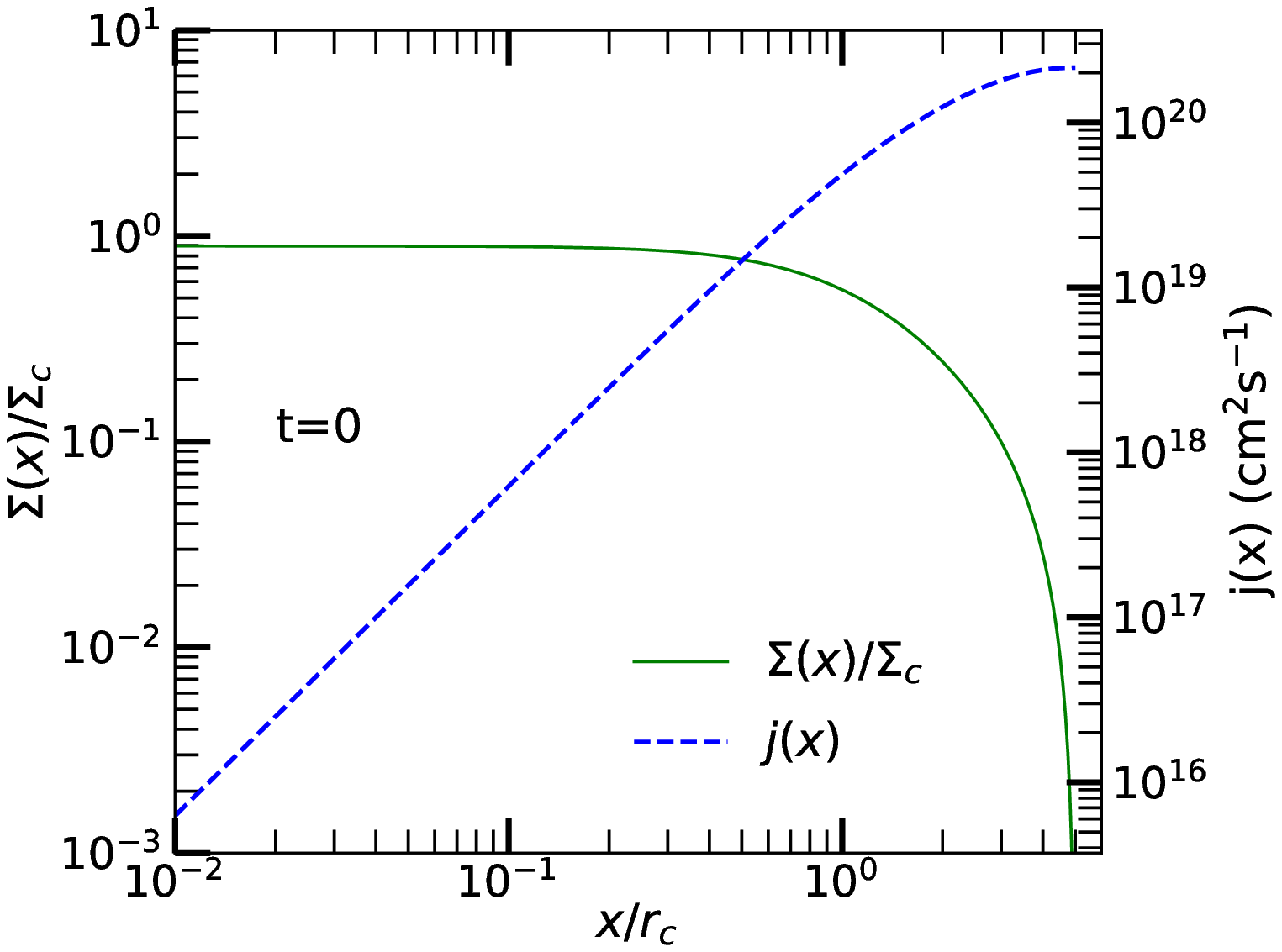}{}
\caption{
Column density $\Sigma(x)/\Sigma_c$ (green), the specific angular momentum $j(x$) (blue) as a function of radial offset $x/r_c$ at $t=0$ (analytic solution).  
}
\label{fig:Rout5_2D_radial_analytic_j_sigma}
\end{figure}
We discuss the semi-analytic approach to evaluate the specific angular momentum $j(x)$ as a function of radial offset $x/r_c$.
We find that the integral is analytically tractable for $t=0$. Inserting Equation (\ref{eq:rho_mod}) into Equation (\ref{eq:sigma_x_formula}), the closed-form expression for the column density is
\begin{equation}
\begin{aligned}
    \tilde{\Sigma}(\tilde{x}) & = \frac{2}{\sqrt{1+\tilde{x}^2}}  
    \left[{\rm arctan}\left(\sqrt{\frac{\tilde{R}_{\rm out}^2 - \tilde{x}^2}{1 +\tilde{x}^2}} \right) \right. \\
    & - \left. \frac{1}{\tilde{R}_{\rm out}} \left(\sqrt{(\tilde{R}_{\rm out}^2 - \tilde{x}^2)(1 +\tilde{x}^2})- {\rm arctan}\left(\sqrt{\frac{\tilde{R}_{\rm out}^2 - \tilde{x}^2}{1 +\tilde{x}^2}}\right) \right) \right] \ ,
\label{eq:sigmax}
\end{aligned}
\end{equation}
where $\tilde{\Sigma} = \Sigma/\Sigma_0$, $\Sigma_0 = \rho_c r_c$, $\tilde{x} = x/r_c$, $\tilde{R}_{\rm out} = R_{\rm out}/r_c$, and the central column density is $\Sigma_c = 2\,\Sigma_0\, {\rm arctan} (\tilde{R}_{\rm out})$; 
some of these parameters are defined earlier in Section \ref{sec:j_numerical}. The enclosed mass is found by integrating Equation (\ref{eq:sigmax}) according to Equation (\ref{eq:M_x_formula}), yielding
\begin{equation}
\begin{aligned}
    \tilde{M}(\tilde{x}) & = 2 \left[\sqrt{1 +\tilde{x}^2} \, {\rm arctan}\left(\sqrt{\frac{\tilde{R}_{\rm out}^2 - \tilde{x}^2}{1 +\tilde{x}^2}} \right) - {\rm arctan}(\tilde{R}_{\rm out}) + \tilde{R}_{\rm out} \right. \\
    & \left. - \ \sqrt{\tilde{R}_{\rm out}^2 - \tilde{x}^2} - \frac{1}{\tilde{R}^2_{\rm out}} \left( \frac{\tilde{R}^3_{\rm out}}{3} - \left(\frac{(\tilde{R}^2_{\rm out} - \tilde{x}^2)^{\frac{3}{2}}}{3}\right) + \sqrt{\tilde{R}_{\rm out}^2 - \tilde{x}^2} \right. \right.\\
    & \left. \left. - \tilde{R}_{\rm out} + \ {\rm arctan} (\tilde{R}_{\rm out}) - \sqrt{1 +\tilde{x}^2} \ {\rm arctan}\left(\sqrt{\frac{\tilde{R}_{\rm out}^2 - \tilde{x}^2}{1 +\tilde{x}^2}} \right)  \right) \right] \ ,
\end{aligned}
\label{eq:Massx}
\end{equation}
where $\tilde{M}=M/(\rho_c r_c^3)$.
We note that the effect of the boundary is exclusively
contained in the factor in square brackets. If the quantity $R_{\rm out}$ is large, the power law profile is more pronounced. However, if $R_{\rm out}/r_c$ is of the order of unity (e.g. for our case it is 5), then the cut off dominates close enough to the flat region to prevent the appearance of the power law. In Figure \ref{fig:Rout5_2D_radial_analytic_j_sigma} the green dashed line shows the analytic $\Sigma/\Sigma_c$ curve, which is flat within the inner region $x/r_c \lesssim 1$, and then falls sharply. The integrated $M(x)$ attains a saturation at the outer edge. Keeping in mind at $t=0$, $M_{\rm tot}(x) = M(x)$. During the pre-collapse phase, $j(x)$ increases linearly with radial offset $x/r_c$ and gets saturated at the very outer radius. This behaviour of $j(x)$ implies the conservation of angular momentum.
\bsp	
\label{lastpage}
\end{document}